\documentclass[aps,twocolumn,english,aip,superscriptaddress,amssymb, showpacs,preprintnumbers,amsmath,amssymb,superscriptaddress]{revtex4-1}

\usepackage{graphicx}
\usepackage{color}
\usepackage{dcolumn}
\usepackage{bm}
\newcommand{\nc}{\newcommand}
\nc{\be}{\begin{equation}}
\nc{\ee}{\end{equation}}
\nc{\bea}{\begin{eqnarray}}
\nc{\eea}{\end{eqnarray}}
\nc{\bean}{\begin{eqnarray*}}
\nc{\eean}{\end{eqnarray*}}
\nc{\mb}{\mbox}
\nc{\rnc}{\renewcommand}
\nc{\vk}{\mb{\bf k}}
\nc{\vp}{\mb{\bf p}}
\nc{\vn}{\mb{\bf n}}
\nc{\vq}{\mb{\bf q}}
\nc{\rr}{\mb{\bf r}}
\nc{\vz}{\hat {\mb{\bf z}}}
\nc{\vj}{\mb{\boldmath$j$}}
\nc{\vg}{\mb{\boldmath$g$}}
\nc{\x}{\mb{\boldmath$x$}}
\nc{\A}{\mb{\boldmath$A$}}
\nc{\va}{\mb{\boldmath$a$}}
\nc{\vs}{\mb{\boldmath$\sigma$}}
\nc{\vpi}{\mb{\boldmath$\pi$}}
\nc{\nab}{\nabla}
\nc{\X}{\sf x}

\begin{document}
\title{Emergent magneto-multipoles and nonlinear responses of a magnetic hopfion}

\author{Yizhou Liu}
\affiliation{RIKEN Center for Emergent Matter Science (CEMS), Wako, Saitama 351-0198, Japan}

\author{Hikaru Watanabe}
\affiliation{RIKEN Center for Emergent Matter Science (CEMS), Wako, Saitama 351-0198, Japan}

\author{Naoto Nagaosa}
\affiliation{RIKEN Center for Emergent Matter Science (CEMS), Wako, Saitama 351-0198, Japan}
\affiliation{Department of Applied Physics, University of Tokyo, Tokyo 113-8656, Japan}

\begin{abstract}
The three-dimensional emergent magnetic field $\textbf{B}^e$ of a magnetic hopfion gives rise to {\it emergent} magneto-multipoles in a similar manner to the multipoles of classical electromagnetic field.
Here, we show that the nonlinear responses of a hopfion are characterized by its emergent magnetic toroidal moment ${T}^e_z = \frac{1}{2}\int (\textbf{r}\times \textbf{B}^e)_z dV$ and emergent magnetic octupole component ${\it \Gamma}^e =\int [(x^2+y^2)B^e_z - xz B^e_x - y z B^e_y] dV$.
%
%
The hopfion exhibits nonreciprocal dynamics (nonlinear hopfion Hall effect) under an ac driving current applied along (perpendicular to) the direction of ${T}^e_z$.
The sign of nonreciprocity and nonlinear Hall angle is determined by the polarity and chirality of hopfion.
The nonlinear electrical transport induced by a magnetic hopfion is also discussed.
This work reveals the vital roles of emergent magneto-multipoles in nonlinear hopfion dynamics and could stimulate further investigations on the dynamical responses of topological spin textures induced by emergent electromagnetic multipoles.
\end{abstract}


\maketitle
Topological spin textures (TSTs) like magnetic skyrmion have sparked intense attentions due to their novel physical properties and attractive potentials for applications~\cite{nagaosa_topological_2013, fert_magnetic_2017}.
Two fundamental quantities characterizing the spin dynamics and the transport phenomena associated with a spin texture are the vector and scalar spin chirality, which are related to the broken spatial-inversion $\mathcal{I}$ and time-reversal symmetry $\mathcal{T}$, respectively.
The scalar spin chirality acts as an effective real-space Berry curvature (the emergent magnetic field $\textbf{B}^e$) on conduction electrons and its spatial integral reflects the topological skyrmion number.
To date, tremendous efforts have been made to study scalar spin chirality related effects, leading to the discovery of many fascinating phenomena such as topological Hall effect~\cite{binz_chirality_2008,neubauer_topological_2009,nagaosa_anomalous_2010}, skyrmion Hall effect~\cite{zang_dynamics_2011, jiang_direct_2017,litzius_skyrmion_2017}, emergent electromagnetic inductance~\cite{nagaosa_emergent_2019,yokouchi_emergent_2020}, etc.
Therefore, identifying such fundamental quantities relevant to the spin texture's dynamics and transport properties is of vital importance.

The multipole originated from classical electromagnetism has been a well-established fundamental concept in many branches of physics~\cite{Landau_classical, thorne_multipole_1980,dubovik_toroid_1990}. 
In condensed matter systems, the multipoles defined within different contents can serve as alluring sources of generating nonlinear and/or nonreciprocal responses~\cite{spaldin_toroidal_2008, tokura_nonreciprocal_2018,arima_resonant_2005, matsuno_ferromagnetic_2007,cheong_broken_2018,bhowal_magnetoelectric_2022}.
%
%
The multipole concept can also be applied to the emergent magnetic field $\textbf{B}^e$~\cite{nagaosa_topological_2013, nagaosa_emergent_2019} associated with TSTs in which {\it emergent} magneto-multipoles may be further identified.
However, unlike their counter part such as the momentum-space Berry curvature dipole and multipole~\cite{sodemann_quantum_2015, ma_observation_2019, zhang_higher_order_2021, du_nonlinear_2021}, the role of emergent magneto-multipoles has not been well explored so far.

%

%
%

Recently, interest in realizing three-dimensional (3D) TSTs in magnetic materials emerges with a focus on their rich 3D topologies and veiled physical properties~\cite{faddeev_stable_1997, three_dimensional_spintronics, donnelly_three_dimensional_2017, donnelly_experimental_2021, yang_chiral_2021}.
One example is the magnetic hopfion, a class of TST with diverse 3D structures~\cite{sutcliffe_skyrmion_2017, liu_binding_2018, sutcliffe_hopfions_2018, tai_static_2018, rybakov_magnetic_2019, voinescu_hopf_2020, kent_creation_2021} characterized by the Hopf charge $Q_H$~\cite{whitehead_expression_1947, wilczek_linking_1983, manton_topological_2004}.
%
%
Even the simplest form of hopfion (i.e., $Q_H=1$) hosts sophisticated spin structure and 3D emergent magnetic field, leading to interesting dynamical properties such as the entangled current-driven  dynamics~\cite{liu_three_dimensional_2020, wang_current_driven_2019} and field-driven resonance modes~\cite{raftrey_field-driven_2021, bo_spin_2021, khodzhaev_hopfion_2022, li_mutual_2022}.
Hence, the 3D structure of $\textbf{B}^e$ of hopfion provides a versatile platform to explore the role of emergent magneto-multipoles.

\begin{figure}
\includegraphics[width=2in,height=1.6in]{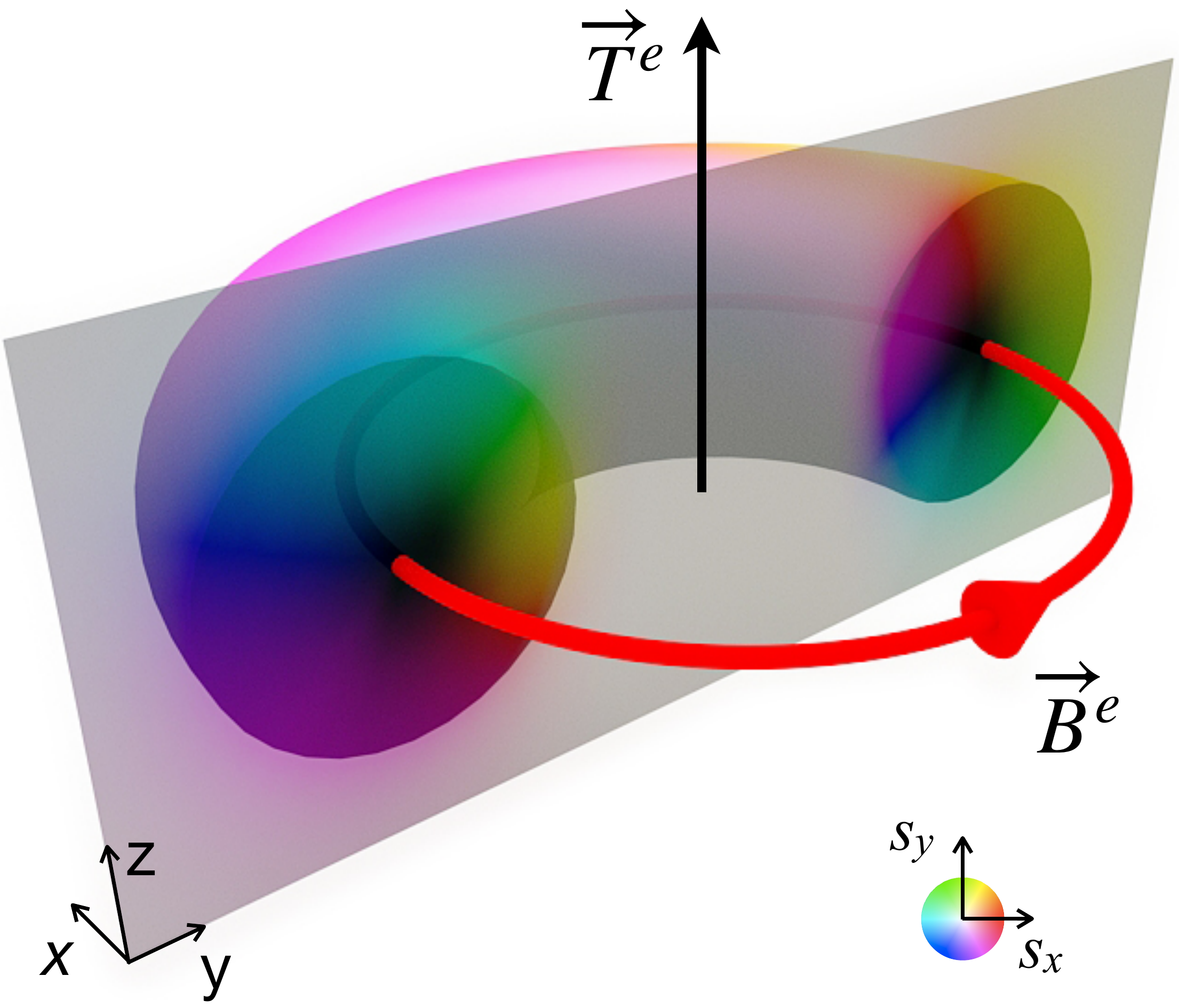}
\caption{Schematic view of a $Q_H = 1$ magnetic hopfion , its emergent magnetic field ($\textbf{B}^e$), and the corresponding emergent toroidal moment ($\textbf{T}^e$).
The half torus represents the equi-spin surface with $S_z=0$.
The rectangle plane shows the cross-sectional view of hopfion onto $yz$ plane, which includes a pair of skyrmion and antiskyrmion.
In the color scheme, black color in the core of torus indicates $S_z=-1$ and white color in the background indicates $S_z=1$. 
The color wheel stands for in-plane spin directions.}
\label{fig:schematic}
\end{figure}

In this Letter, we show that the emergent magneto-multipoles of a $Q_H = 1$ magnetic hopfion characterizes the nonlinear responses of both the spin dynamics and electron transports.
The torus-like hopfion possesses swirling-like $\textbf{B}^e$ and it gives rise to an emergent toroidal moment $\textbf{T}^e$ like the classical toroidal moment of a torus solenoid~\cite{dubovik_toroid_1990, gobel_topological_2020, pershoguba_electronic_2021}.
It is found that the hopfion shows nonlinear nonreciprocal translation and rotation with respect to the direction of $\textbf{T}^e$ under an ac electrical current applied along $\textbf{T}^e$.
When the ac current is applied perpendicular to $\textbf{T}^e$, a nonlinear hopfion Hall effect is identified where the hopfion also translates and rotates with respect to $\textbf{T}^e$.
These nonlinear dynamics of hopfion is governed by the emergent toroidal moment ${T}^e_z$  and the emergent magnetic octupole component $\it \Gamma^e$.
On the other hand, we also demonstrate that the hopfion gives rise to nonreciprocal magnetoresistance and nonlinear Hall effect in its transport properties of conduction electrons.
%
%
From these aspects, the emergent magneto-multipole serves as a new fundamental quantity that characterizes the nonlinear responses of spin textures.

The typical spin configuration of a hopfion $\textbf{S}_0(\textbf{r})$ with $Q_H = 1$ is shown in Fig.~\ref{fig:schematic}.
The Hopf charge is defined as $Q_H = \int \textbf{A}^e(\textbf{r})\cdot \textbf{B}^e(\textbf{r}) dV$, where $B^e_i(\textbf{r})=\frac{1}{2} \varepsilon_{ijk} \textbf{S} \cdot (\partial_j \textbf{S} \times \partial_k \textbf{S})$ is the emergent magnetic field and $\textbf{A}^e$ is the corresponding vector potential ($\textbf{B}^e=\nabla \times \textbf{A}^e$).
This hopfion structure is axially symmetric around $z$-axis.
The associated emergent magnetic field exhibits a swirling like structure as indicated by the red looped arrow in Fig.~\ref{fig:schematic}
Although the symmetry of hopfion results in a vanishing spatial-averaged emergent magnetic field $<\textbf{B}^e (\textbf{r})>=0$, it defines a finite {emergent toroidal moment} $\textbf{T}^e = \frac{1}{2} \int \textbf{r} \times \textbf{B}^e (\textbf{r}) dV = T_z^e \hat{z}$ along the hopfion's symmetry axis ($z$-axis)~\cite{gobel_topological_2020, pershoguba_electronic_2021}.
Note that $\textbf{T}^e$ appears as the integral of the Berry connection over the spin texture (see Eqs.(S13) and (S14) in the Supplemental Materials~\cite{supp}).
As a result, nonreciprocal responses related to $\textbf{T}^e$ are expected for the hopfion.

Previous study suggests that the current-driven dynamics of hopfion are effectively described by three dynamical modes: translation, rotation, and dilation~\cite{liu_three_dimensional_2020}.
For clarity, here we mainly focus on the dynamics of hopfion related to $z$-axis, which is most relevant to the nonreciprocal responses.
At position ${\bf r} = (x, y, z)$ and time $t$, the translation and rotation mode of hopfion are expressed as ${\textbf S}(\textbf{r},t) = {\textbf S}_0 (\hat{O}(\textbf{r} - \textbf{Z}))$, where {\bf{Z}}=$Z \hat{z}$ characterizes the translation of hopfion along $z$-axis, and $\hat{O}$ is the rotation operator.
At infinitesimal rotation, for only the rotation around $z$-axis, we have $\hat{O}  \approx  1 - i{{\theta}}_z {{L}}_z$, where $\hat{L_{i}} = -i\varepsilon_{ijk} r_j \partial_{k}$ is the angular momentum operator and $\theta_z$ is the rotation angle of hopfion around $z$-axis.
The dilation of hopfion can be expressed as $\textbf{S}(\textbf{r},t) = \textbf{S}_0(\lambda \textbf{r}) $ and the static equilibrium dilation $\lambda_0=1$ at the initial state.

Based on these dynamical modes, a glance at the dynamics of hopfion along its symmetry axis can be deduced from the spin's Berry phase term of the Lagrangian~\cite{auerbach_interacting_1994, tatara_microscopic_2008, zang_dynamics_2011}
$L_{BP} = \int (1 - \text{cos}\theta) \dot{\phi} dV$,
where $\theta$ and $\phi$ are the polar and azimuthal angle of the localized spin $\textbf{S}$ with unit length.
The variation of the spin's Berry phase term $\delta L_{BP} = \int \textbf{S} \cdot \delta \textbf{S}  \times \dot{\textbf{S}} dV$ can be expressed in terms of the slowly-varying dynamical modes of hopfion after integrating out the spin configuration,
\begin{equation} 
\label{Eq:Berry_dynamics_Z}
\delta L^z_{BP} = \it (-T^e_z \dot{Z} + \Gamma^e \dot{\theta}_z) \delta \lambda,
\end{equation}
where $T^e_z$ is the emergent toroidal moment and ${\it \Gamma^e} = \int [(x^2+y^2)B^e_z - xz B^e_x - y z B^e_y] dV $ is an emergent magnetic octupole component of hopfion ~\cite{watanabe_group_theoretical_2018,hayami_classification_2018}.
In Eq.(\ref{Eq:Berry_dynamics_Z}), the translation and rotation are coupled to the dilation via $T^e_z$ and $\it \Gamma^e$ rather than the emergent magnetic field, whose spatial average vanishes for the hopfion.
This distinguishes the dynamics of hopfion from other previously studied TSTs in which the dynamical modes (e.g., longitudinal and transverse motion) are coupled via the gyrovector (total emergent magnetic flux of the spin texture)~\cite{papanicolaou_dynamics_1991,guslienko_magnetic_2008, tretiakov_dynamics_2008,zang_dynamics_2011,everschor_sitte_real_space_2014, nagaosa_topological_2013,psaroudaki_skyrmions_2018}.
In principle, the velocity coupled with toroidal moment can generate nonreciprocal effects~\cite{cheong_broken_2018, cheong_magnetic_2022}, and thus the nonreciprocal dynamics of hopfion is also expected from the coupled dynamical modes.

\begin{figure}
\begin{center}
\includegraphics[width=3.2in,height=3.5in]{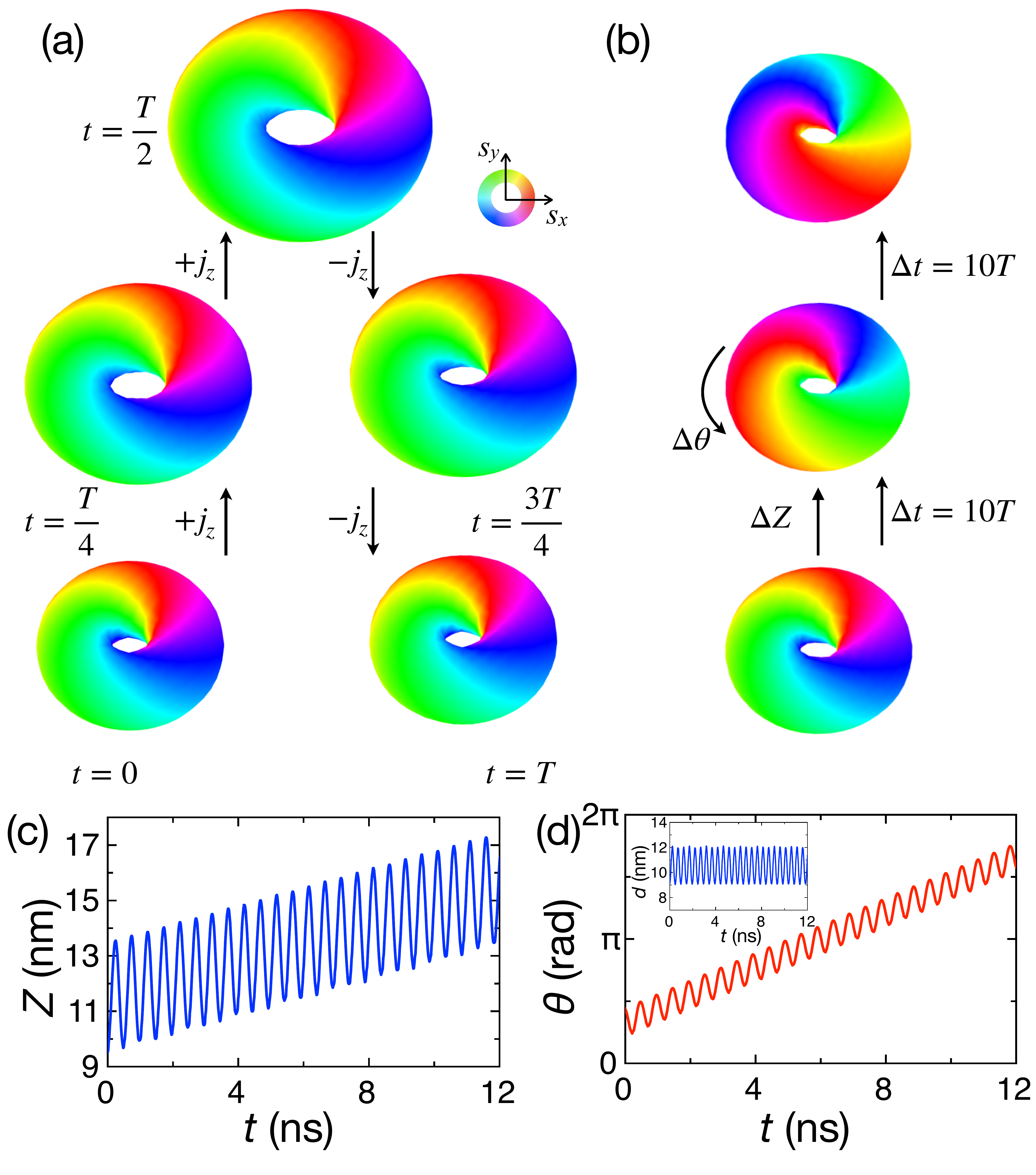}
\caption{Nonreciprocal dynamics of a magnetic hopfion driven by a sinusoidal-type ac current applied along $z$-axis.
(a) Snapshots of the hopfion dynamics within a single driving cycle.
(b) Snapshots of the hopfion dynamics showing clear net translation and rotation after 10 and 20 driving cycles.
(c) and (d) show the translation and rotation of the ac hopfion dynamics.
Inset of (d) also shows the change of hopfion's size over time.
}
\label{fig:snapshot}
\end{center}
\end{figure}

To carefully check the nonreciprocal dynamics of hopfion, spin dynamics simulations were performed.
A magnetic hopfion stabilized in a frustrated magnet is employed as the initial state.
The corresponding model Hamiltonian is $\mathcal{H} =  - \sum\limits_{<i,j>} J_{ij}  \textbf{S}_i \cdot  \textbf{S}_j$, where the summation of the exchange interaction is extended up to fourth nearest neighbor on a cubic mesh~\cite{supp}.
The presence of hopfion spontaneously breaks the inversion symmetry of this system.

The spin dynamics are calculated by solving the Landau-Lifshitz-Gilbert (LLG) equation with the spin transfer torque (STT) terms~\cite{slonczewski_current-driven_1996, berger_emission_1996, zhang_roles_2004}:
\begin{equation} 
\label{Eq:LLG}
\begin{split}
\frac{d\textbf{S}}{dt} = -\gamma_0 \textbf{S} \times \textbf{H}_{\text{eff}} + \frac{\alpha}{S} \textbf{S} \times &\frac{d\textbf{S}}{dt}  + \frac{p a^3}{2eS} (\textbf{j}\cdot \bold{\nabla}) \textbf{S}\\
&+ \frac{p a^3 \beta}{2eS^2} (\textbf{j}\cdot \bold{\nabla})  \textbf{S} \times \textbf{S}.
\end{split}
\end{equation}
Here $\gamma_0$ is the gyromagnetic ratio, $\alpha$ is the damping constant, $p$ is the spin polarization, $a$ is the lattice constant, and $\textbf{H}_{\text{eff}}$ is the effective magnetic field. 
$S$ is the spin length, which is fixed to be 1 for simplicity.
The effects of STT are described by the last two terms in Eq.~\ref{Eq:LLG} with the current applied along the hopfion's symmetry axis ($z$-axis).
$\textbf{j}$ is the current density and $\beta$ quantifies the non-adiabaticity of STT.

\begin{figure}
\includegraphics[width=3.in,height=2.125in]{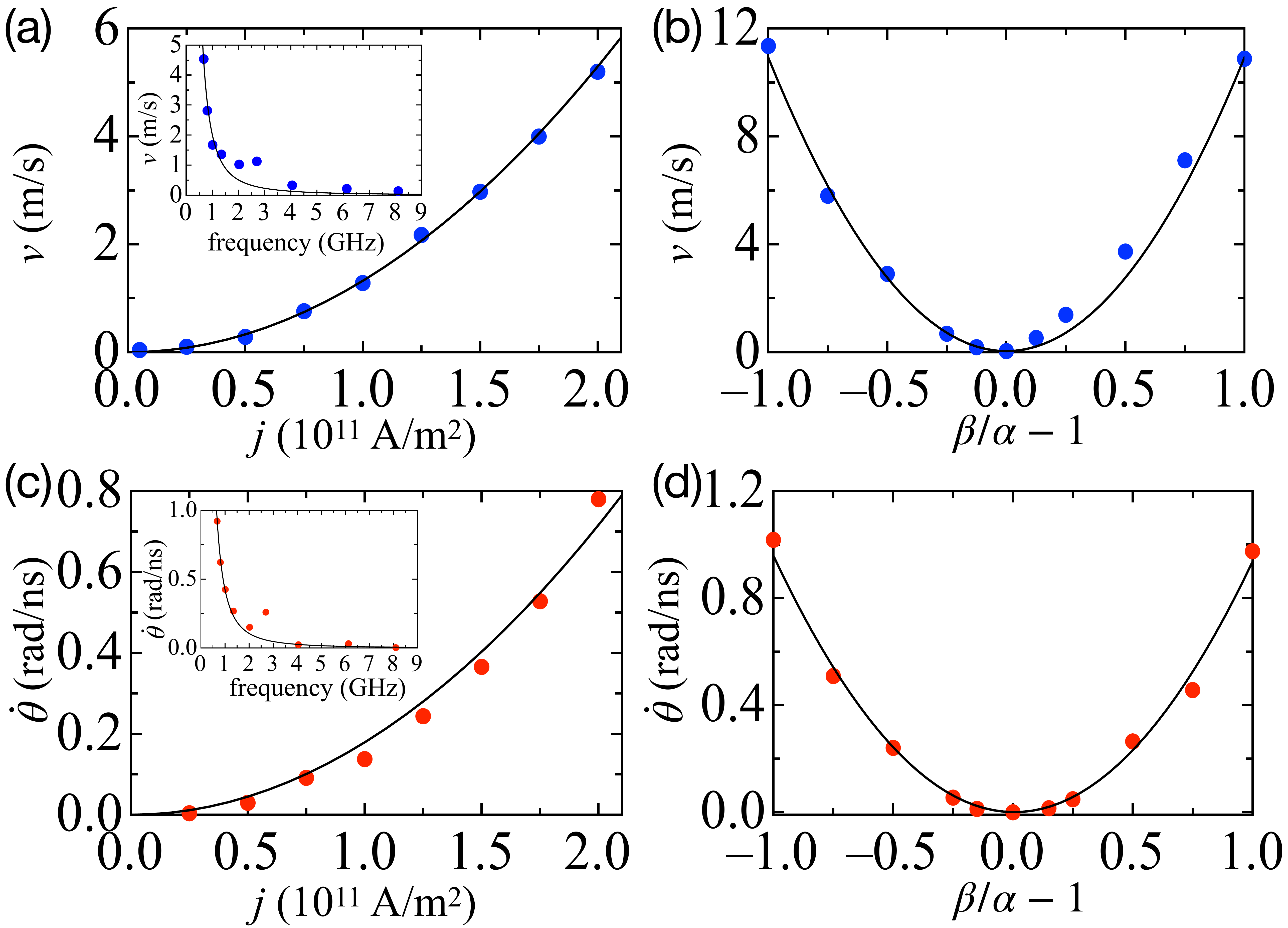}
\caption{Current density and ($\beta/\alpha-1$) dependence of hopfion's net velocity [(a) and (b)] and net angular velocity [(c) and (d)].
Inset in (a) and (c) is the corresponding driving frequency dependence. 
Blue and red dots are simulation results. 
Black lines represent the $j^2$ fitting in (a) and (c), the $\frac{1}{\omega^2}$ fitting in the inset of (a) and (c), and the $(\beta / \alpha-1)^2$ fitting in (b) and (d), respectively.
The ac current is applied along $z$-axis.
}
\label{fig:dependence}
\end{figure}

Typical results of the hopfion dynamics under an ac driving current $j_z = j \text{sin}(\omega t)$ applied along $z-$axis are summarized in Fig.~\ref{fig:snapshot}.
As shown in the snapshots [Fig.~\ref{fig:snapshot}(a)] within a single driving period, for the first (second) half of the period with positive (negative) applied current, the hopfion moves along the current direction with a clockwise (counterclockwise) rotation and an expansion (shrinkage).
Interestingly, after each cycle, the hopfion gains a net translation and rotation.
Snapshots after multiple cycles with more pronounced net translation and rotation are shown in Fig.~\ref{fig:snapshot}(b) for clarity (also see movie for the nonreciprocal dynamics in the Supplemental Materials~\cite{supp}). 
The steady-state ac dynamics plotted in Fig.~\ref{fig:snapshot} (c) and (d) also show that a net translation and rotation are associated with the oscillating dynamics and they increase with time in a linear manner.
Therefore, the hopfion possesses nonreciprocal translation and rotation under an ac driving current.
On the other hand, the size of hopfion $d$ oscillates around an equilibrium value [Fig.~\ref{fig:snapshot}(d) inset] so that the steady-state ac dynamics can be sustained.
The current density dependences of hopfion's net velocity $v$ and net angular velocity $\dot{\theta}$ are plotted in Fig.~\ref{fig:dependence}(a) and (c).
The quadratic dependence on the current density suggests a nonlinear nonreciprocal dynamics of hopfion.
%


\begin{figure}
\begin{center}
\includegraphics[width=3.5in,height=1.3in]{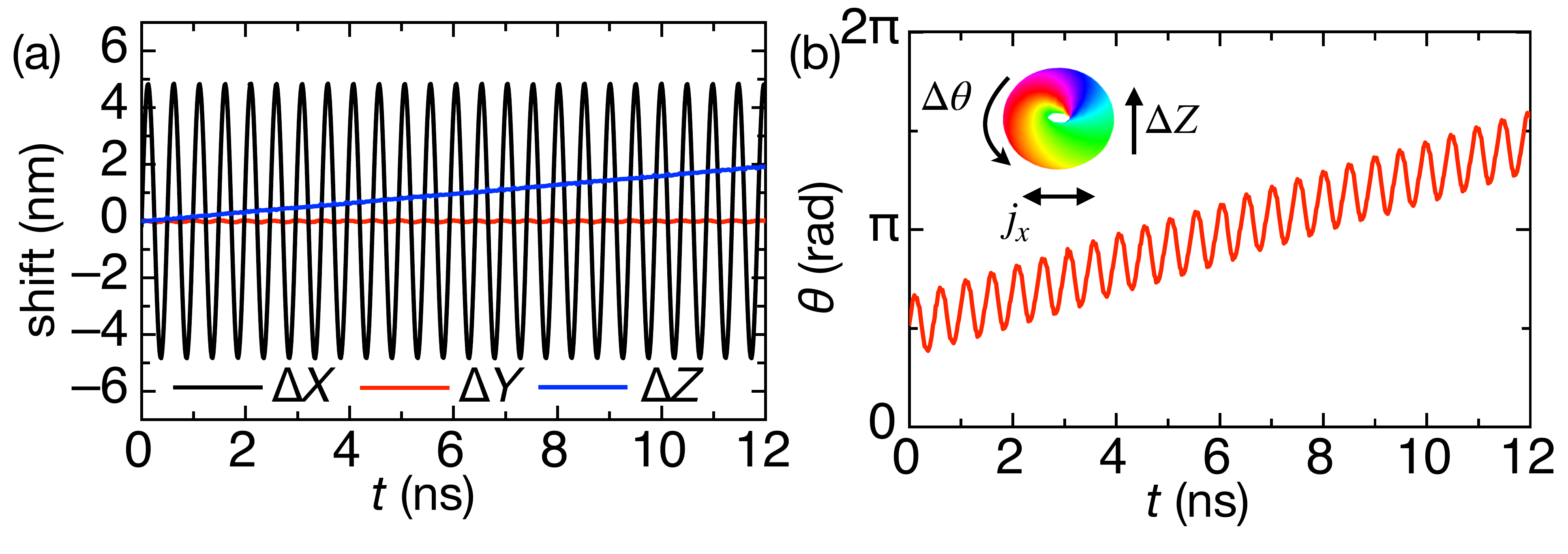}
\caption{Nonlinear hopfion Hall effect under an ac current applied along $x$-axis.
(a) The translation of ac hopfion dynamics along different axes.
(b) The ac rotation dynamics of hopfion around $z-$axis.
Inset of (b) shows the schematic of the nonlinear hopfion Hall effect.}
\label{fig:inplane}
\end{center}
\end{figure}

To better understand this nonlinear nonreciprocal dynamics, we derive the corresponding equation of motion via the generalized Thiele's approach based on the dynamical modes of hopfion~\cite{thiele_steady-state_1973, tretiakov_dynamics_2008,liu_three_dimensional_2020}.
Other than the translation, rotation, and dilation of hopfion, it is necessary to take into account two additional effects in order to capture the nonreciprocal dynamics.
Firstly, there is always an energy change associated with the dilation of hopfion.
Although this energy change can be ignored in describing the linear dc dynamics, it is quite essential here for generating the nonlinear nonreciprocal dynamics.
The formation of hopfion requires a minimum number of spins and it assigns an energy limit bounded with the shrinking of hopfion.
Hence, the energy of hopfion is asymmetric with respect to the dilation (expansion and shrinkage) and can be effectively described by an anharmonic potential $E= \frac{1}{2}m (\lambda-1)^2 + \frac{1}{6} \chi (\lambda-1)^3$ (for more details, see Supplemental Materials~\cite{supp}).
In addition, the dilation also leads to an effective field $\textbf{H}_{\lambda} = \frac{\zeta}{\mu_s} \textbf{S}_0 \times [(\lambda -1) \textbf{r} \cdot \partial_{\textbf{r}}]\textbf{S}_0$ acted on the hopfion, where $\mu_s$ is the magnetic moment and $\zeta$ is a phenomenological parameter similar to $m$ and $\gamma$.
These two factors associated with the dilation of hopfion are key ingredients for understanding the nonreciprocal dynamics of hopfion beyond the linear response regime.

Based on these dynamical modes, the equations of motion for hopfion are obtained as
\begin{align}
\label{Eq1:OOP0}
-{T^e_z} \dot{\lambda} + \alpha K_{RR} \dot{Z} + \alpha K_{R\it \Theta} \dot{ \it \theta_z} =& \beta K_{RR}\xi j_z -\frac{\gamma_0 \zeta}{\mu_s}T^e_z (\lambda-1),\\
\label{Eq1:OOP1}
{\it \Gamma^e} \dot{\lambda} + \alpha K_{R\it \Theta} \dot{Z} + \alpha K_{\it \Theta \Theta} \dot{ \it \theta_z} =& \beta K_{R \Theta}\xi j_z +\frac{\gamma_0 \zeta}{\mu_s}{\it \Gamma^e} (\lambda-1),\\
\label{Eq1:OOP2}
\nonumber T^e_z \dot{Z} -{\it \Gamma^e}  \dot{ \it \theta_z} + \alpha K_{\lambda \lambda} \dot{\lambda} = \xi T^e_z j_z&- \frac{\gamma_0 }{\mu_s}m(\lambda-1)\\
&\text{           }- \frac{\gamma_0}{2\mu_s} \chi (\lambda -1)^2,
\end{align}
with $K_{RR} = \int (\partial_z \textbf{S})^2 dV$, $K_{R\it \Theta} = \int \partial_z \textbf{S} \cdot (x\partial_y - y \partial_x) \textbf{S} dV$, $K_{\it \Theta \Theta} = \int [(x\partial_y - y \partial_x) \textbf{S}]^2 dV$, and $K_{\lambda \lambda} = \int (\textbf{r} \cdot \partial_{\textbf{r}} \textbf{S})^2 dV$.
By setting the current density $j_z=j \text{sin}(\omega t)$, these equations can be solved in the ac limit  hierarchically with respect to the order of $j$.

We then solve the dynamical modes as $\dot{Z} \approx \dot{Z}_1(j) + \dot{Z}_2(j^2) $,  $\dot{\theta_z} \approx \dot{\theta_z}_1(j) + \dot{\theta_z}_2(j^2) $, and $\lambda \approx \lambda_0 +  \lambda_1(j) + \lambda_2(j^2)$ up to the second order of $j$ (for details of the solution, see Supplemental Materials~\cite{supp}).
While the first order solutions (with $j$ linear terms) only contain oscillating dynamics, their second order solutions (with $j^2$ terms) include non-oscillating terms, i.e., the terms related to the nonreciprocal dynamics.
By further imposing the symmetry of hopfion, the following simplified dependences can be obtained~\cite{supp}
%
%
%
\begin{equation}
\begin{split}
\label{Eq:Z2}
\dot{Z}^{dc}_2 \sim \frac{1}{\omega^2} T^e_z (\frac{\beta}{\alpha}-1)^2 j^2,
\end{split}
\end{equation}
\begin{equation}
\begin{split}
\label{Eq:theta2}
\dot{\theta}^{dc}_{z2} \sim -\frac{1}{\omega^2} {\it \Gamma^e} (\frac{\beta}{\alpha}-1)^2 j^2.
\end{split}
\end{equation}
$\dot{Z}_2^{dc}$ and $\dot{\theta}_{z2}^{dc}$ represent the nonlinear nonreciprocal translation and rotation of hopfion.
They have the same dependence on the current density $j$, driving frequency $\omega$, and $(\frac{\beta}{\alpha}-1)$.
These dependences are also consistent with the simulation results as shown in Fig.~\ref{fig:dependence}.
The anomaly in the frequency dependence might be due to some resonances or other internal modes of hopfion, which are difficult to capture using the current formalism.

\begin{figure}
\includegraphics[width=3.4in,height=1.4in]{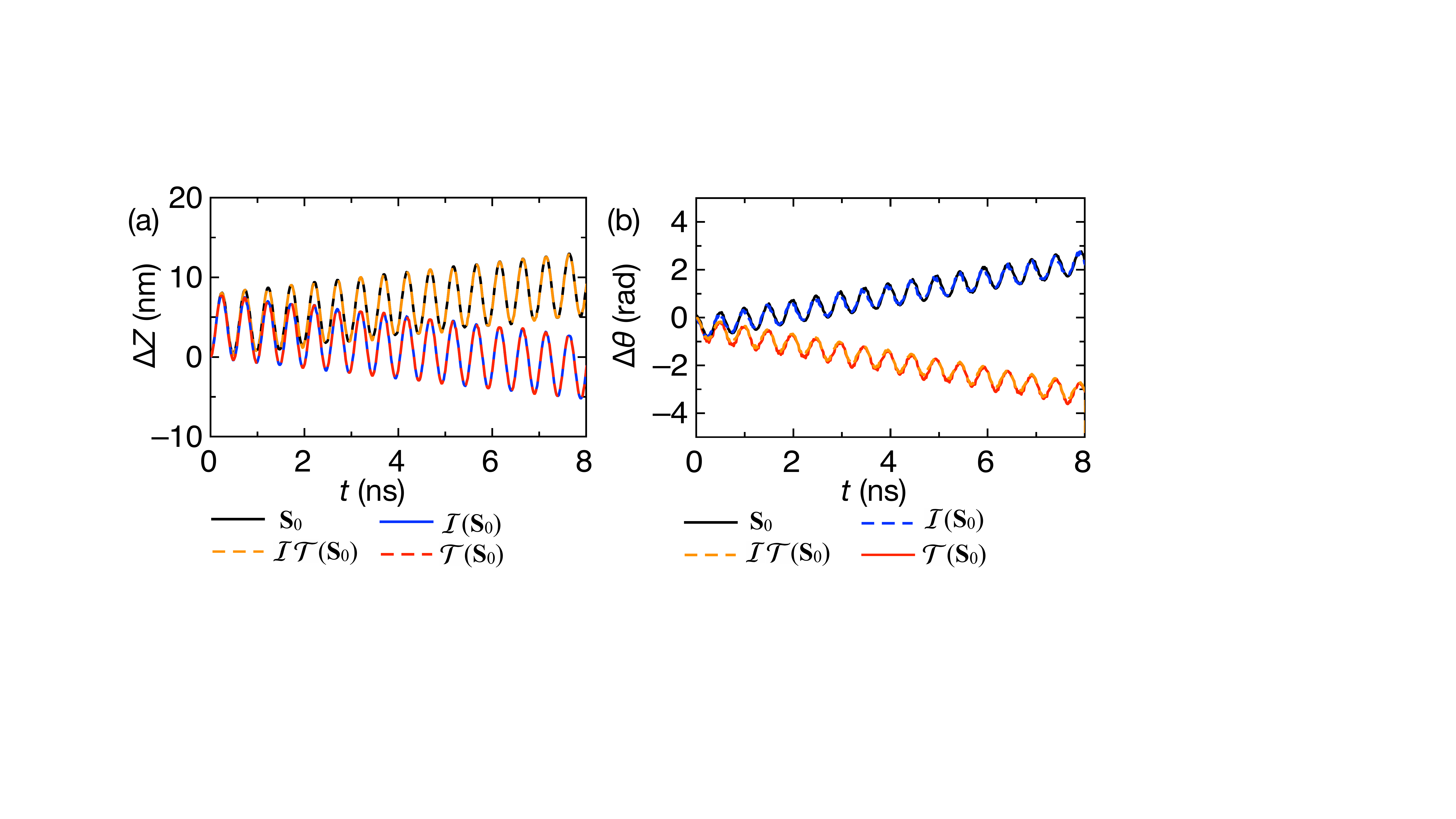}
\caption{Nonreciprocal translation (a) and rotation (b) for different symmetry operations on the hopfion configuration.
%
%
}
\label{fig:symmetry}
\end{figure}

Next, we discuss the nonlinear hopfion responses under an ac current applied perpendicular to the direction of $T^e_z$ (e.g., along $x-$axis).
Previous study suggests the dc hopfion dynamics under an in-plane current include a rotation of the hopfion plane, which is equivalent to a rotation of $T^e_z$~\cite{liu_three_dimensional_2020}.
Therefore, although the current and $T^e_z$ are initially perpendicular to each other, as time increases, the current component along $T^e_z$ develops and nonlinear dynamics of hopfion are still expected.

The simulation results for an ac current applied along $x-$axis are shown in Fig.~\ref{fig:inplane}.
While the hopfion oscillates along $x-$axis, the net translation and rotation are still bounded with $z-$axis.
Hence, the hopfion moves perpendicular to the applied ac current direction, which is a nonlinear hopfion Hall effect.
Moreover, the current density dependence, frequency dependence, and $\frac{\beta}{\alpha}$ dependence of hopfion's net velocity and angular velocity have also been checked, which all agree well with the analytical results (see Supplemental Materials~\cite{supp}).

The equation of motion (Eq.~\ref{Eq:Z2} and Eq.~\ref{Eq:theta2}) indicates that $T^e_z$ and $\it \Gamma^e$ characterize the nonlinear dynamics of hopfion.
The nonreciprocity of $\dot{Z}_2^{dc}$ and $\dot{\theta}_{z2}^{dc}$ are determined by $T^e_z$ and $\it \Gamma^e$, respectively.
Therefore, it is possible to control the nonlinear dynamics of hopfion by manipulating its spin configuration (e.g., switch the chirality or polarity).
Although both the translation and rotation of hopfion show nonlinear responses, the corresponding symmetries are not the same.
$T^e_z$ breaks both $\mathcal{T}$ and $\mathcal{I}$.
As a result, the velocity of hopfion reverses its sign upon $\mathcal{T}$ and $\mathcal{I}$ operation but is invariant under $\mathcal{IT}$ operation [Fig.~\ref{fig:symmetry}(a)].
On the other hand, the rotation of hopfion is determined by $\it \Gamma^e$, a parity-even emergent magneto-multipole.
So the net rotation of hopfion reverses its sign upon $\mathcal{T}$ or $\mathcal{IT}$ only [Fig.~\ref{fig:symmetry}(b)].

Finally, to facilitate a more complete understanding, we study the nonlinear electron transport effects induced by a hopfion, which is effectively the counterpart of the STT effect.
The second-order current density in response to the external electric field $\textbf{E}$ is defined as 
\begin{equation} 
\begin{aligned}
j_{2,a} = \chi_{abc} E_b E_c
\end{aligned}
\end{equation} 
where $\chi_{abc}$ is the second-order conductivity tensor.
By considering a double exchange model (in which the itinerant electrons are coupled to the localized spins via the exchange coupling)~\cite{ye_berry_1999,bruno_topological_2004}, $\chi_{abc}$ can be obtained by solving the semiclassical Boltzmann equation~\cite{isobe_high-frequency_2020, ishizuka_anomalous_2020,isobe_toroidal_2022} (for details of the derivation, see~\cite{supp}).
For the hopfion spin texture studied here ($\textbf{T}^e = T^e_z \hat{\textbf{z}}$), three typical second-order conductivity components $\chi_{zzz}/2 =   \chi_{zxx} =  \chi_{zyy}  = \chi \propto T^e_z$ are obtained (detailed solutions can be found in~\cite{supp}).
So the emergent toroidal moment of hopfion induces two types of nonlinear responses.
$\chi_{zxx}$ and $\chi_{zyy}$ characterize a nonlinear Hall effect and $\chi_{zzz}$ characterizes a nonreciprocal transport both along the direction of $T^e_z$.
These nonlinear transport effects are consistent with the nonlinear spin dynamics of hopfion.

Our results reveal the vital role of the emergent magneto-multipoles in determining the nonlinear spin dynamics and nonlinear electrical transports related to a magnetic hopfion.
We anticipate the hopfions in other physical systems may also show nonlinear responses in a similar manner as the vector field itself breaks at least the inversion symmetry~\cite{volovik_1977, babaev_dual_2002, kawaguchi_knots_2008, ackerman_static_2017}.
As an example, we demonstrate the nonlinear nonreciprocal dynamics of a hopfion stabilized in typical chiral magnet with conical background (see Supplemental Materials~\cite{supp}).
In addition to hopfion, the emergent magneto-multipoles can also be applied to other TSTs in both two- and three-dimensional systems and can be engineered by tailoring different kinds of TSTs.
This work also provides an alternative working principle for the potential implementations of hopfion in 3D spintronic devices~\cite{three_dimensional_spintronics, yang_chiral_2021}

{\em Acknowledgement}---
We thank W. Koshibae for fruitful discussions.
N.N. was supported by JST CREST grant number JMPJCR1874 and JPMJCR16F1, and JSPS KAKENHI grant number 18H03676.
H.W. acknowledged the support from JSPS KAKENHI No. JP21J00453.
Y.L. acknowledges financial support from the RIKEN Special Postdoctoral Researcher (SPDR) Program and JST CREST grant number JPMJCR20T1.


\begin{thebibliography}{70}%
\makeatletter
\providecommand \@ifxundefined [1]{%
 \@ifx{#1\undefined}
}%
\providecommand \@ifnum [1]{%
 \ifnum #1\expandafter \@firstoftwo
 \else \expandafter \@secondoftwo
 \fi
}%
\providecommand \@ifx [1]{%
 \ifx #1\expandafter \@firstoftwo
 \else \expandafter \@secondoftwo
 \fi
}%
\providecommand \natexlab [1]{#1}%
\providecommand \enquote  [1]{``#1''}%
\providecommand \bibnamefont  [1]{#1}%
\providecommand \bibfnamefont [1]{#1}%
\providecommand \citenamefont [1]{#1}%
\providecommand \href@noop [0]{\@secondoftwo}%
\providecommand \href [0]{\begingroup \@sanitize@url \@href}%
\providecommand \@href[1]{\@@startlink{#1}\@@href}%
\providecommand \@@href[1]{\endgroup#1\@@endlink}%
\providecommand \@sanitize@url [0]{\catcode `\\12\catcode `\$12\catcode
  `\&12\catcode `\#12\catcode `\^12\catcode `\_12\catcode `\%12\relax}%
\providecommand \@@startlink[1]{}%
\providecommand \@@endlink[0]{}%
\providecommand \url  [0]{\begingroup\@sanitize@url \@url }%
\providecommand \@url [1]{\endgroup\@href {#1}{\urlprefix }}%
\providecommand \urlprefix  [0]{URL }%
\providecommand \Eprint [0]{\href }%
\providecommand \doibase [0]{http://dx.doi.org/}%
\providecommand \selectlanguage [0]{\@gobble}%
\providecommand \bibinfo  [0]{\@secondoftwo}%
\providecommand \bibfield  [0]{\@secondoftwo}%
\providecommand \translation [1]{[#1]}%
\providecommand \BibitemOpen [0]{}%
\providecommand \bibitemStop [0]{}%
\providecommand \bibitemNoStop [0]{.\EOS\space}%
\providecommand \EOS [0]{\spacefactor3000\relax}%
\providecommand \BibitemShut  [1]{\csname bibitem#1\endcsname}%
\let\auto@bib@innerbib\@empty
\bibitem [{\citenamefont {Nagaosa}\ and\ \citenamefont
  {Tokura}(2013)}]{nagaosa_topological_2013}%
  \BibitemOpen
  \bibfield  {author} {\bibinfo {author} {\bibfnamefont {N.}~\bibnamefont
  {Nagaosa}}\ and\ \bibinfo {author} {\bibfnamefont {Y.}~\bibnamefont
  {Tokura}},\ }\href {\doibase 10.1038/nnano.2013.243} {\bibfield  {journal}
  {\bibinfo  {journal} {Nature Nanotech}\ }\textbf {\bibinfo {volume} {8}},\
  \bibinfo {pages} {899} (\bibinfo {year} {2013})}\BibitemShut {NoStop}%
\bibitem [{\citenamefont {Fert}\ \emph {et~al.}(2017)\citenamefont {Fert},
  \citenamefont {Reyren},\ and\ \citenamefont {Cros}}]{fert_magnetic_2017}%
  \BibitemOpen
  \bibfield  {author} {\bibinfo {author} {\bibfnamefont {A.}~\bibnamefont
  {Fert}}, \bibinfo {author} {\bibfnamefont {N.}~\bibnamefont {Reyren}}, \ and\
  \bibinfo {author} {\bibfnamefont {V.}~\bibnamefont {Cros}},\ }\href {\doibase
  10.1038/natrevmats.2017.31} {\bibfield  {journal} {\bibinfo  {journal} {Nat
  Rev Mater}\ }\textbf {\bibinfo {volume} {2}},\ \bibinfo {pages} {1} (\bibinfo
  {year} {2017})}\BibitemShut {NoStop}%
\bibitem [{\citenamefont {Binz}\ and\ \citenamefont
  {Vishwanath}(2008)}]{binz_chirality_2008}%
  \BibitemOpen
  \bibfield  {author} {\bibinfo {author} {\bibfnamefont {B.}~\bibnamefont
  {Binz}}\ and\ \bibinfo {author} {\bibfnamefont {A.}~\bibnamefont
  {Vishwanath}},\ }\href {\doibase 10.1016/j.physb.2007.10.136} {\bibfield
  {journal} {\bibinfo  {journal} {Physica B: Condensed Matter}\ }\textbf
  {\bibinfo {volume} {403}},\ \bibinfo {pages} {1336} (\bibinfo {year}
  {2008})}\BibitemShut {NoStop}%
\bibitem [{\citenamefont {Neubauer}\ \emph {et~al.}(2009)\citenamefont
  {Neubauer}, \citenamefont {Pfleiderer}, \citenamefont {Binz}, \citenamefont
  {Rosch}, \citenamefont {Ritz}, \citenamefont {Niklowitz},\ and\ \citenamefont
  {Böni}}]{neubauer_topological_2009}%
  \BibitemOpen
  \bibfield  {author} {\bibinfo {author} {\bibfnamefont {A.}~\bibnamefont
  {Neubauer}}, \bibinfo {author} {\bibfnamefont {C.}~\bibnamefont
  {Pfleiderer}}, \bibinfo {author} {\bibfnamefont {B.}~\bibnamefont {Binz}},
  \bibinfo {author} {\bibfnamefont {A.}~\bibnamefont {Rosch}}, \bibinfo
  {author} {\bibfnamefont {R.}~\bibnamefont {Ritz}}, \bibinfo {author}
  {\bibfnamefont {P.~G.}\ \bibnamefont {Niklowitz}}, \ and\ \bibinfo {author}
  {\bibfnamefont {P.}~\bibnamefont {Böni}},\ }\href {\doibase
  10.1103/PhysRevLett.102.186602} {\bibfield  {journal} {\bibinfo  {journal}
  {Phys. Rev. Lett.}\ }\textbf {\bibinfo {volume} {102}},\ \bibinfo {pages}
  {186602} (\bibinfo {year} {2009})}\BibitemShut {NoStop}%
\bibitem [{\citenamefont {Nagaosa}\ \emph {et~al.}(2010)\citenamefont
  {Nagaosa}, \citenamefont {Sinova}, \citenamefont {Onoda}, \citenamefont
  {MacDonald},\ and\ \citenamefont {Ong}}]{nagaosa_anomalous_2010}%
  \BibitemOpen
  \bibfield  {author} {\bibinfo {author} {\bibfnamefont {N.}~\bibnamefont
  {Nagaosa}}, \bibinfo {author} {\bibfnamefont {J.}~\bibnamefont {Sinova}},
  \bibinfo {author} {\bibfnamefont {S.}~\bibnamefont {Onoda}}, \bibinfo
  {author} {\bibfnamefont {A.~H.}\ \bibnamefont {MacDonald}}, \ and\ \bibinfo
  {author} {\bibfnamefont {N.~P.}\ \bibnamefont {Ong}},\ }\href {\doibase
  10.1103/RevModPhys.82.1539} {\bibfield  {journal} {\bibinfo  {journal} {Rev.
  Mod. Phys.}\ }\textbf {\bibinfo {volume} {82}},\ \bibinfo {pages} {1539}
  (\bibinfo {year} {2010})}\BibitemShut {NoStop}%
\bibitem [{\citenamefont {Zang}\ \emph {et~al.}(2011)\citenamefont {Zang},
  \citenamefont {Mostovoy}, \citenamefont {Han},\ and\ \citenamefont
  {Nagaosa}}]{zang_dynamics_2011}%
  \BibitemOpen
  \bibfield  {author} {\bibinfo {author} {\bibfnamefont {J.}~\bibnamefont
  {Zang}}, \bibinfo {author} {\bibfnamefont {M.}~\bibnamefont {Mostovoy}},
  \bibinfo {author} {\bibfnamefont {J.~H.}\ \bibnamefont {Han}}, \ and\
  \bibinfo {author} {\bibfnamefont {N.}~\bibnamefont {Nagaosa}},\ }\href
  {\doibase 10.1103/PhysRevLett.107.136804} {\bibfield  {journal} {\bibinfo
  {journal} {Phys. Rev. Lett.}\ }\textbf {\bibinfo {volume} {107}},\ \bibinfo
  {pages} {136804} (\bibinfo {year} {2011})}\BibitemShut {NoStop}%
\bibitem [{\citenamefont {Jiang}\ \emph {et~al.}(2017)\citenamefont {Jiang},
  \citenamefont {Zhang}, \citenamefont {Yu}, \citenamefont {Zhang},
  \citenamefont {Wang}, \citenamefont {Benjamin~Jungfleisch}, \citenamefont
  {Pearson}, \citenamefont {Cheng}, \citenamefont {Heinonen}, \citenamefont
  {Wang}, \citenamefont {Zhou}, \citenamefont {Hoffmann},\ and\ \citenamefont
  {te~Velthuis}}]{jiang_direct_2017}%
  \BibitemOpen
  \bibfield  {author} {\bibinfo {author} {\bibfnamefont {W.}~\bibnamefont
  {Jiang}}, \bibinfo {author} {\bibfnamefont {X.}~\bibnamefont {Zhang}},
  \bibinfo {author} {\bibfnamefont {G.}~\bibnamefont {Yu}}, \bibinfo {author}
  {\bibfnamefont {W.}~\bibnamefont {Zhang}}, \bibinfo {author} {\bibfnamefont
  {X.}~\bibnamefont {Wang}}, \bibinfo {author} {\bibfnamefont {M.}~\bibnamefont
  {Benjamin~Jungfleisch}}, \bibinfo {author} {\bibfnamefont {J.~E.}\
  \bibnamefont {Pearson}}, \bibinfo {author} {\bibfnamefont {X.}~\bibnamefont
  {Cheng}}, \bibinfo {author} {\bibfnamefont {O.}~\bibnamefont {Heinonen}},
  \bibinfo {author} {\bibfnamefont {K.~L.}\ \bibnamefont {Wang}}, \bibinfo
  {author} {\bibfnamefont {Y.}~\bibnamefont {Zhou}}, \bibinfo {author}
  {\bibfnamefont {A.}~\bibnamefont {Hoffmann}}, \ and\ \bibinfo {author}
  {\bibfnamefont {S.~G.~E.}\ \bibnamefont {te~Velthuis}},\ }\href {\doibase
  10.1038/nphys3883} {\bibfield  {journal} {\bibinfo  {journal} {Nature
  Physics}\ }\textbf {\bibinfo {volume} {13}},\ \bibinfo {pages} {162}
  (\bibinfo {year} {2017})}\BibitemShut {NoStop}%
\bibitem [{\citenamefont {Litzius}\ \emph {et~al.}(2017)\citenamefont
  {Litzius}, \citenamefont {Lemesh}, \citenamefont {Kr{\"u}ger}, \citenamefont
  {Bassirian}, \citenamefont {Caretta}, \citenamefont {Richter}, \citenamefont
  {B{\"u}ttner}, \citenamefont {Sato}, \citenamefont {Tretiakov}, \citenamefont
  {F{\"o}rster}, \citenamefont {Reeve}, \citenamefont {Weigand}, \citenamefont
  {Bykova}, \citenamefont {Stoll}, \citenamefont {Sch{\"u}tz}, \citenamefont
  {Beach},\ and\ \citenamefont {Kl{\"a}ui}}]{litzius_skyrmion_2017}%
  \BibitemOpen
  \bibfield  {author} {\bibinfo {author} {\bibfnamefont {K.}~\bibnamefont
  {Litzius}}, \bibinfo {author} {\bibfnamefont {I.}~\bibnamefont {Lemesh}},
  \bibinfo {author} {\bibfnamefont {B.}~\bibnamefont {Kr{\"u}ger}}, \bibinfo
  {author} {\bibfnamefont {P.}~\bibnamefont {Bassirian}}, \bibinfo {author}
  {\bibfnamefont {L.}~\bibnamefont {Caretta}}, \bibinfo {author} {\bibfnamefont
  {K.}~\bibnamefont {Richter}}, \bibinfo {author} {\bibfnamefont
  {F.}~\bibnamefont {B{\"u}ttner}}, \bibinfo {author} {\bibfnamefont
  {K.}~\bibnamefont {Sato}}, \bibinfo {author} {\bibfnamefont {O.~A.}\
  \bibnamefont {Tretiakov}}, \bibinfo {author} {\bibfnamefont {J.}~\bibnamefont
  {F{\"o}rster}}, \bibinfo {author} {\bibfnamefont {R.~M.}\ \bibnamefont
  {Reeve}}, \bibinfo {author} {\bibfnamefont {M.}~\bibnamefont {Weigand}},
  \bibinfo {author} {\bibfnamefont {I.}~\bibnamefont {Bykova}}, \bibinfo
  {author} {\bibfnamefont {H.}~\bibnamefont {Stoll}}, \bibinfo {author}
  {\bibfnamefont {G.}~\bibnamefont {Sch{\"u}tz}}, \bibinfo {author}
  {\bibfnamefont {G.~S.~D.}\ \bibnamefont {Beach}}, \ and\ \bibinfo {author}
  {\bibfnamefont {M.}~\bibnamefont {Kl{\"a}ui}},\ }\href {\doibase
  10.1038/nphys4000} {\bibfield  {journal} {\bibinfo  {journal} {Nature
  Physics}\ }\textbf {\bibinfo {volume} {13}},\ \bibinfo {pages} {170}
  (\bibinfo {year} {2017})}\BibitemShut {NoStop}%
\bibitem [{\citenamefont {Nagaosa}(2019)}]{nagaosa_emergent_2019}%
  \BibitemOpen
  \bibfield  {author} {\bibinfo {author} {\bibfnamefont {N.}~\bibnamefont
  {Nagaosa}},\ }\href {\doibase 10.2183/pjab.95.019} {\bibfield  {journal}
  {\bibinfo  {journal} {Proceedings of the Japan Academy, Series B}\ }\textbf
  {\bibinfo {volume} {95}},\ \bibinfo {pages} {278} (\bibinfo {year}
  {2019})}\BibitemShut {NoStop}%
\bibitem [{\citenamefont {Yokouchi}\ \emph {et~al.}(2020)\citenamefont
  {Yokouchi}, \citenamefont {Kagawa}, \citenamefont {Hirschberger},
  \citenamefont {Otani}, \citenamefont {Nagaosa},\ and\ \citenamefont
  {Tokura}}]{yokouchi_emergent_2020}%
  \BibitemOpen
  \bibfield  {author} {\bibinfo {author} {\bibfnamefont {T.}~\bibnamefont
  {Yokouchi}}, \bibinfo {author} {\bibfnamefont {F.}~\bibnamefont {Kagawa}},
  \bibinfo {author} {\bibfnamefont {M.}~\bibnamefont {Hirschberger}}, \bibinfo
  {author} {\bibfnamefont {Y.}~\bibnamefont {Otani}}, \bibinfo {author}
  {\bibfnamefont {N.}~\bibnamefont {Nagaosa}}, \ and\ \bibinfo {author}
  {\bibfnamefont {Y.}~\bibnamefont {Tokura}},\ }\href {\doibase
  10.1038/s41586-020-2775-x} {\bibfield  {journal} {\bibinfo  {journal}
  {Nature}\ }\textbf {\bibinfo {volume} {586}},\ \bibinfo {pages} {232}
  (\bibinfo {year} {2020})}\BibitemShut {NoStop}%
\bibitem [{\citenamefont {Landau}\ and\ \citenamefont
  {Lifshitz}(1975)}]{Landau_classical}%
  \BibitemOpen
  \bibfield  {author} {\bibinfo {author} {\bibfnamefont {L.~D.}\ \bibnamefont
  {Landau}}\ and\ \bibinfo {author} {\bibfnamefont {E.~M.}\ \bibnamefont
  {Lifshitz}},\ }\href {\doibase 10.1007/978-1-4612-0869-3} {\emph {\bibinfo
  {title} {The {Classical} {Theory} of {Fields} - 4th {Edition}}}},\ Graduate
  {Texts} in {Contemporary} {Physics}\ (\bibinfo  {publisher}
  {(Butterworth-Heinemann},\ \bibinfo {address} {London},\ \bibinfo {year}
  {1975})\BibitemShut {NoStop}%
\bibitem [{\citenamefont {Thorne}(1980)}]{thorne_multipole_1980}%
  \BibitemOpen
  \bibfield  {author} {\bibinfo {author} {\bibfnamefont {K.~S.}\ \bibnamefont
  {Thorne}},\ }\href {\doibase 10.1103/RevModPhys.52.299} {\bibfield  {journal}
  {\bibinfo  {journal} {Rev. Mod. Phys.}\ }\textbf {\bibinfo {volume} {52}},\
  \bibinfo {pages} {299} (\bibinfo {year} {1980})}\BibitemShut {NoStop}%
\bibitem [{\citenamefont {Dubovik}\ and\ \citenamefont
  {Tugushev}(1990)}]{dubovik_toroid_1990}%
  \BibitemOpen
  \bibfield  {author} {\bibinfo {author} {\bibfnamefont {V.~M.}\ \bibnamefont
  {Dubovik}}\ and\ \bibinfo {author} {\bibfnamefont {V.~V.}\ \bibnamefont
  {Tugushev}},\ }\href {\doibase 10.1016/0370-1573(90)90042-Z} {\bibfield
  {journal} {\bibinfo  {journal} {Physics Reports}\ }\textbf {\bibinfo {volume}
  {187}},\ \bibinfo {pages} {145} (\bibinfo {year} {1990})}\BibitemShut
  {NoStop}%
\bibitem [{\citenamefont {Spaldin}\ \emph {et~al.}(2008)\citenamefont
  {Spaldin}, \citenamefont {Fiebig},\ and\ \citenamefont
  {Mostovoy}}]{spaldin_toroidal_2008}%
  \BibitemOpen
  \bibfield  {author} {\bibinfo {author} {\bibfnamefont {N.~A.}\ \bibnamefont
  {Spaldin}}, \bibinfo {author} {\bibfnamefont {M.}~\bibnamefont {Fiebig}}, \
  and\ \bibinfo {author} {\bibfnamefont {M.}~\bibnamefont {Mostovoy}},\ }\href
  {\doibase 10.1088/0953-8984/20/43/434203} {\bibfield  {journal} {\bibinfo
  {journal} {J. Phys.: Condens. Matter}\ }\textbf {\bibinfo {volume} {20}},\
  \bibinfo {pages} {434203} (\bibinfo {year} {2008})}\BibitemShut {NoStop}%
\bibitem [{\citenamefont {Tokura}\ and\ \citenamefont
  {Nagaosa}(2018)}]{tokura_nonreciprocal_2018}%
  \BibitemOpen
  \bibfield  {author} {\bibinfo {author} {\bibfnamefont {Y.}~\bibnamefont
  {Tokura}}\ and\ \bibinfo {author} {\bibfnamefont {N.}~\bibnamefont
  {Nagaosa}},\ }\href {\doibase 10.1038/s41467-018-05759-4} {\bibfield
  {journal} {\bibinfo  {journal} {Nat Commun}\ }\textbf {\bibinfo {volume}
  {9}},\ \bibinfo {pages} {3740} (\bibinfo {year} {2018})}\BibitemShut
  {NoStop}%
\bibitem [{\citenamefont {Arima}\ \emph {et~al.}(2005)\citenamefont {Arima},
  \citenamefont {Jung}, \citenamefont {Matsubara}, \citenamefont {Kubota},
  \citenamefont {He}, \citenamefont {Kaneko},\ and\ \citenamefont
  {Tokura}}]{arima_resonant_2005}%
  \BibitemOpen
  \bibfield  {author} {\bibinfo {author} {\bibfnamefont {T.-h.}\ \bibnamefont
  {Arima}}, \bibinfo {author} {\bibfnamefont {J.-H.}\ \bibnamefont {Jung}},
  \bibinfo {author} {\bibfnamefont {M.}~\bibnamefont {Matsubara}}, \bibinfo
  {author} {\bibfnamefont {M.}~\bibnamefont {Kubota}}, \bibinfo {author}
  {\bibfnamefont {J.-P.}\ \bibnamefont {He}}, \bibinfo {author} {\bibfnamefont
  {Y.}~\bibnamefont {Kaneko}}, \ and\ \bibinfo {author} {\bibfnamefont
  {Y.}~\bibnamefont {Tokura}},\ }\href {\doibase 10.1143/JPSJ.74.1419}
  {\bibfield  {journal} {\bibinfo  {journal} {J. Phys. Soc. Jpn.}\ }\textbf
  {\bibinfo {volume} {74}},\ \bibinfo {pages} {1419} (\bibinfo {year}
  {2005})}\BibitemShut {NoStop}%
\bibitem [{\citenamefont {Matsuno}\ \emph {et~al.}(2007)\citenamefont
  {Matsuno}, \citenamefont {Lottermoser}, \citenamefont {Arima}, \citenamefont
  {Kawasaki},\ and\ \citenamefont {Tokura}}]{matsuno_ferromagnetic_2007}%
  \BibitemOpen
  \bibfield  {author} {\bibinfo {author} {\bibfnamefont {J.}~\bibnamefont
  {Matsuno}}, \bibinfo {author} {\bibfnamefont {T.}~\bibnamefont
  {Lottermoser}}, \bibinfo {author} {\bibfnamefont {T.}~\bibnamefont {Arima}},
  \bibinfo {author} {\bibfnamefont {M.}~\bibnamefont {Kawasaki}}, \ and\
  \bibinfo {author} {\bibfnamefont {Y.}~\bibnamefont {Tokura}},\ }\href
  {\doibase 10.1103/PhysRevB.75.180403} {\bibfield  {journal} {\bibinfo
  {journal} {Phys. Rev. B}\ }\textbf {\bibinfo {volume} {75}},\ \bibinfo
  {pages} {180403} (\bibinfo {year} {2007})}\BibitemShut {NoStop}%
\bibitem [{\citenamefont {Cheong}\ \emph {et~al.}(2018)\citenamefont {Cheong},
  \citenamefont {Talbayev}, \citenamefont {Kiryukhin},\ and\ \citenamefont
  {Saxena}}]{cheong_broken_2018}%
  \BibitemOpen
  \bibfield  {author} {\bibinfo {author} {\bibfnamefont {S.-W.}\ \bibnamefont
  {Cheong}}, \bibinfo {author} {\bibfnamefont {D.}~\bibnamefont {Talbayev}},
  \bibinfo {author} {\bibfnamefont {V.}~\bibnamefont {Kiryukhin}}, \ and\
  \bibinfo {author} {\bibfnamefont {A.}~\bibnamefont {Saxena}},\ }\href
  {\doibase 10.1038/s41535-018-0092-5} {\bibfield  {journal} {\bibinfo
  {journal} {npj Quant Mater}\ }\textbf {\bibinfo {volume} {3}},\ \bibinfo
  {pages} {1} (\bibinfo {year} {2018})}\BibitemShut {NoStop}%
\bibitem [{\citenamefont {Bhowal}\ and\ \citenamefont
  {Spaldin}(2022)}]{bhowal_magnetoelectric_2022}%
  \BibitemOpen
  \bibfield  {author} {\bibinfo {author} {\bibfnamefont {S.}~\bibnamefont
  {Bhowal}}\ and\ \bibinfo {author} {\bibfnamefont {N.~A.}\ \bibnamefont
  {Spaldin}},\ }\href {\doibase 10.1103/PhysRevLett.128.227204} {\bibfield
  {journal} {\bibinfo  {journal} {Phys. Rev. Lett.}\ }\textbf {\bibinfo
  {volume} {128}},\ \bibinfo {pages} {227204} (\bibinfo {year}
  {2022})}\BibitemShut {NoStop}%
\bibitem [{\citenamefont {Sodemann}\ and\ \citenamefont
  {Fu}(2015)}]{sodemann_quantum_2015}%
  \BibitemOpen
  \bibfield  {author} {\bibinfo {author} {\bibfnamefont {I.}~\bibnamefont
  {Sodemann}}\ and\ \bibinfo {author} {\bibfnamefont {L.}~\bibnamefont {Fu}},\
  }\href {\doibase 10.1103/PhysRevLett.115.216806} {\bibfield  {journal}
  {\bibinfo  {journal} {Phys. Rev. Lett.}\ }\textbf {\bibinfo {volume} {115}},\
  \bibinfo {pages} {216806} (\bibinfo {year} {2015})}\BibitemShut {NoStop}%
\bibitem [{\citenamefont {Ma}\ \emph {et~al.}(2019)\citenamefont {Ma},
  \citenamefont {Xu}, \citenamefont {Shen}, \citenamefont {MacNeill},
  \citenamefont {Fatemi}, \citenamefont {Chang}, \citenamefont {Mier~Valdivia},
  \citenamefont {Wu}, \citenamefont {Du}, \citenamefont {Hsu}, \citenamefont
  {Fang}, \citenamefont {Gibson}, \citenamefont {Watanabe}, \citenamefont
  {Taniguchi}, \citenamefont {Cava}, \citenamefont {Kaxiras}, \citenamefont
  {Lu}, \citenamefont {Lin}, \citenamefont {Fu}, \citenamefont {Gedik},\ and\
  \citenamefont {Jarillo-Herrero}}]{ma_observation_2019}%
  \BibitemOpen
  \bibfield  {author} {\bibinfo {author} {\bibfnamefont {Q.}~\bibnamefont
  {Ma}}, \bibinfo {author} {\bibfnamefont {S.-Y.}\ \bibnamefont {Xu}}, \bibinfo
  {author} {\bibfnamefont {H.}~\bibnamefont {Shen}}, \bibinfo {author}
  {\bibfnamefont {D.}~\bibnamefont {MacNeill}}, \bibinfo {author}
  {\bibfnamefont {V.}~\bibnamefont {Fatemi}}, \bibinfo {author} {\bibfnamefont
  {T.-R.}\ \bibnamefont {Chang}}, \bibinfo {author} {\bibfnamefont {A.~M.}\
  \bibnamefont {Mier~Valdivia}}, \bibinfo {author} {\bibfnamefont
  {S.}~\bibnamefont {Wu}}, \bibinfo {author} {\bibfnamefont {Z.}~\bibnamefont
  {Du}}, \bibinfo {author} {\bibfnamefont {C.-H.}\ \bibnamefont {Hsu}},
  \bibinfo {author} {\bibfnamefont {S.}~\bibnamefont {Fang}}, \bibinfo {author}
  {\bibfnamefont {Q.~D.}\ \bibnamefont {Gibson}}, \bibinfo {author}
  {\bibfnamefont {K.}~\bibnamefont {Watanabe}}, \bibinfo {author}
  {\bibfnamefont {T.}~\bibnamefont {Taniguchi}}, \bibinfo {author}
  {\bibfnamefont {R.~J.}\ \bibnamefont {Cava}}, \bibinfo {author}
  {\bibfnamefont {E.}~\bibnamefont {Kaxiras}}, \bibinfo {author} {\bibfnamefont
  {H.-Z.}\ \bibnamefont {Lu}}, \bibinfo {author} {\bibfnamefont
  {H.}~\bibnamefont {Lin}}, \bibinfo {author} {\bibfnamefont {L.}~\bibnamefont
  {Fu}}, \bibinfo {author} {\bibfnamefont {N.}~\bibnamefont {Gedik}}, \ and\
  \bibinfo {author} {\bibfnamefont {P.}~\bibnamefont {Jarillo-Herrero}},\
  }\href {\doibase 10.1038/s41586-018-0807-6} {\bibfield  {journal} {\bibinfo
  {journal} {Nature}\ }\textbf {\bibinfo {volume} {565}},\ \bibinfo {pages}
  {337} (\bibinfo {year} {2019})}\BibitemShut {NoStop}%
\bibitem [{\citenamefont {Zhang}\ \emph {et~al.}(2021)\citenamefont {Zhang},
  \citenamefont {Gao}, \citenamefont {Xie}, \citenamefont {Po},\ and\
  \citenamefont {Law}}]{zhang_higher_order_2021}%
  \BibitemOpen
  \bibfield  {author} {\bibinfo {author} {\bibfnamefont {C.-P.}\ \bibnamefont
  {Zhang}}, \bibinfo {author} {\bibfnamefont {X.-J.}\ \bibnamefont {Gao}},
  \bibinfo {author} {\bibfnamefont {Y.-M.}\ \bibnamefont {Xie}}, \bibinfo
  {author} {\bibfnamefont {H.~C.}\ \bibnamefont {Po}}, \ and\ \bibinfo {author}
  {\bibfnamefont {K.~T.}\ \bibnamefont {Law}},\ }\href
  {http://arxiv.org/abs/2012.15628} {\bibfield  {journal} {\bibinfo  {journal}
  {arXiv:2012.15628}\ } (\bibinfo {year} {2021})}\BibitemShut {NoStop}%
\bibitem [{\citenamefont {Du}\ \emph {et~al.}(2021)\citenamefont {Du},
  \citenamefont {Lu},\ and\ \citenamefont {Xie}}]{du_nonlinear_2021}%
  \BibitemOpen
  \bibfield  {author} {\bibinfo {author} {\bibfnamefont {Z.~Z.}\ \bibnamefont
  {Du}}, \bibinfo {author} {\bibfnamefont {H.-Z.}\ \bibnamefont {Lu}}, \ and\
  \bibinfo {author} {\bibfnamefont {X.~C.}\ \bibnamefont {Xie}},\ }\href
  {\doibase 10.1038/s42254-021-00359-6} {\bibfield  {journal} {\bibinfo
  {journal} {Nat. Rev. Phys.}\ }\textbf {\bibinfo {volume} {3}},\ \bibinfo
  {pages} {744} (\bibinfo {year} {2021})}\BibitemShut {NoStop}%
\bibitem [{\citenamefont {Faddeev}\ and\ \citenamefont
  {Niemi}(1997)}]{faddeev_stable_1997}%
  \BibitemOpen
  \bibfield  {author} {\bibinfo {author} {\bibfnamefont {L.}~\bibnamefont
  {Faddeev}}\ and\ \bibinfo {author} {\bibfnamefont {A.~J.}\ \bibnamefont
  {Niemi}},\ }\href {\doibase 10.1038/387058a0} {\bibfield  {journal} {\bibinfo
   {journal} {Nature}\ }\textbf {\bibinfo {volume} {387}},\ \bibinfo {pages}
  {58} (\bibinfo {year} {1997})}\BibitemShut {NoStop}%
\bibitem [{\citenamefont {Fern{\'a}ndez-Pacheco}\ \emph
  {et~al.}(2017)\citenamefont {Fern{\'a}ndez-Pacheco}, \citenamefont
  {Streubel}, \citenamefont {Fruchart}, \citenamefont {Hertel}, \citenamefont
  {Fischer},\ and\ \citenamefont {Cowburn}}]{three_dimensional_spintronics}%
  \BibitemOpen
  \bibfield  {author} {\bibinfo {author} {\bibfnamefont {A.}~\bibnamefont
  {Fern{\'a}ndez-Pacheco}}, \bibinfo {author} {\bibfnamefont {R.}~\bibnamefont
  {Streubel}}, \bibinfo {author} {\bibfnamefont {O.}~\bibnamefont {Fruchart}},
  \bibinfo {author} {\bibfnamefont {R.}~\bibnamefont {Hertel}}, \bibinfo
  {author} {\bibfnamefont {P.}~\bibnamefont {Fischer}}, \ and\ \bibinfo
  {author} {\bibfnamefont {R.~P.}\ \bibnamefont {Cowburn}},\ }\href {\doibase
  10.1038/ncomms15756} {\bibfield  {journal} {\bibinfo  {journal} {Nature
  Communications}\ }\textbf {\bibinfo {volume} {8}},\ \bibinfo {pages} {15756}
  (\bibinfo {year} {2017})}\BibitemShut {NoStop}%
\bibitem [{\citenamefont {Donnelly}\ \emph {et~al.}(2017)\citenamefont
  {Donnelly}, \citenamefont {Guizar-Sicairos}, \citenamefont {Scagnoli},
  \citenamefont {Gliga}, \citenamefont {Holler}, \citenamefont {Raabe},\ and\
  \citenamefont {Heyderman}}]{donnelly_three_dimensional_2017}%
  \BibitemOpen
  \bibfield  {author} {\bibinfo {author} {\bibfnamefont {C.}~\bibnamefont
  {Donnelly}}, \bibinfo {author} {\bibfnamefont {M.}~\bibnamefont
  {Guizar-Sicairos}}, \bibinfo {author} {\bibfnamefont {V.}~\bibnamefont
  {Scagnoli}}, \bibinfo {author} {\bibfnamefont {S.}~\bibnamefont {Gliga}},
  \bibinfo {author} {\bibfnamefont {M.}~\bibnamefont {Holler}}, \bibinfo
  {author} {\bibfnamefont {J.}~\bibnamefont {Raabe}}, \ and\ \bibinfo {author}
  {\bibfnamefont {L.~J.}\ \bibnamefont {Heyderman}},\ }\href {\doibase
  10.1038/nature23006} {\bibfield  {journal} {\bibinfo  {journal} {Nature}\
  }\textbf {\bibinfo {volume} {547}},\ \bibinfo {pages} {328} (\bibinfo {year}
  {2017})}\BibitemShut {NoStop}%
\bibitem [{\citenamefont {Donnelly}\ \emph {et~al.}(2021)\citenamefont
  {Donnelly}, \citenamefont {Metlov}, \citenamefont {Scagnoli}, \citenamefont
  {Guizar-Sicairos}, \citenamefont {Holler}, \citenamefont {Bingham},
  \citenamefont {Raabe}, \citenamefont {Heyderman}, \citenamefont {Cooper},\
  and\ \citenamefont {Gliga}}]{donnelly_experimental_2021}%
  \BibitemOpen
  \bibfield  {author} {\bibinfo {author} {\bibfnamefont {C.}~\bibnamefont
  {Donnelly}}, \bibinfo {author} {\bibfnamefont {K.~L.}\ \bibnamefont
  {Metlov}}, \bibinfo {author} {\bibfnamefont {V.}~\bibnamefont {Scagnoli}},
  \bibinfo {author} {\bibfnamefont {M.}~\bibnamefont {Guizar-Sicairos}},
  \bibinfo {author} {\bibfnamefont {M.}~\bibnamefont {Holler}}, \bibinfo
  {author} {\bibfnamefont {N.~S.}\ \bibnamefont {Bingham}}, \bibinfo {author}
  {\bibfnamefont {J.}~\bibnamefont {Raabe}}, \bibinfo {author} {\bibfnamefont
  {L.~J.}\ \bibnamefont {Heyderman}}, \bibinfo {author} {\bibfnamefont {N.~R.}\
  \bibnamefont {Cooper}}, \ and\ \bibinfo {author} {\bibfnamefont
  {S.}~\bibnamefont {Gliga}},\ }\href {\doibase 10.1038/s41567-020-01057-3}
  {\bibfield  {journal} {\bibinfo  {journal} {Nat. Phys.}\ }\textbf {\bibinfo
  {volume} {17}},\ \bibinfo {pages} {316} (\bibinfo {year} {2021})}\BibitemShut
  {NoStop}%
\bibitem [{\citenamefont {Yang}\ \emph {et~al.}(2021)\citenamefont {Yang},
  \citenamefont {Naaman}, \citenamefont {Paltiel},\ and\ \citenamefont
  {Parkin}}]{yang_chiral_2021}%
  \BibitemOpen
  \bibfield  {author} {\bibinfo {author} {\bibfnamefont {S.-H.}\ \bibnamefont
  {Yang}}, \bibinfo {author} {\bibfnamefont {R.}~\bibnamefont {Naaman}},
  \bibinfo {author} {\bibfnamefont {Y.}~\bibnamefont {Paltiel}}, \ and\
  \bibinfo {author} {\bibfnamefont {S.~S.~P.}\ \bibnamefont {Parkin}},\ }\href
  {\doibase 10.1038/s42254-021-00302-9} {\bibfield  {journal} {\bibinfo
  {journal} {Nat. Rev. Phys.}\ }\textbf {\bibinfo {volume} {3}},\ \bibinfo
  {pages} {328} (\bibinfo {year} {2021})}\BibitemShut {NoStop}%
\bibitem [{\citenamefont {Sutcliffe}(2017)}]{sutcliffe_skyrmion_2017}%
  \BibitemOpen
  \bibfield  {author} {\bibinfo {author} {\bibfnamefont {P.}~\bibnamefont
  {Sutcliffe}},\ }\href {\doibase 10.1103/PhysRevLett.118.247203} {\bibfield
  {journal} {\bibinfo  {journal} {Phys. Rev. Lett.}\ }\textbf {\bibinfo
  {volume} {118}},\ \bibinfo {pages} {247203} (\bibinfo {year}
  {2017})}\BibitemShut {NoStop}%
\bibitem [{\citenamefont {Liu}\ \emph {et~al.}(2018)\citenamefont {Liu},
  \citenamefont {Lake},\ and\ \citenamefont {Zang}}]{liu_binding_2018}%
  \BibitemOpen
  \bibfield  {author} {\bibinfo {author} {\bibfnamefont {Y.}~\bibnamefont
  {Liu}}, \bibinfo {author} {\bibfnamefont {R.~K.}\ \bibnamefont {Lake}}, \
  and\ \bibinfo {author} {\bibfnamefont {J.}~\bibnamefont {Zang}},\ }\href
  {\doibase 10.1103/PhysRevB.98.174437} {\bibfield  {journal} {\bibinfo
  {journal} {Phys. Rev. B}\ }\textbf {\bibinfo {volume} {98}},\ \bibinfo
  {pages} {174437} (\bibinfo {year} {2018})}\BibitemShut {NoStop}%
\bibitem [{\citenamefont {Sutcliffe}(2018)}]{sutcliffe_hopfions_2018}%
  \BibitemOpen
  \bibfield  {author} {\bibinfo {author} {\bibfnamefont {P.}~\bibnamefont
  {Sutcliffe}},\ }\href {\doibase 10.1088/1751-8121/aad521} {\bibfield
  {journal} {\bibinfo  {journal} {J. Phys. A: Math. Theor.}\ }\textbf {\bibinfo
  {volume} {51}},\ \bibinfo {pages} {375401} (\bibinfo {year}
  {2018})}\BibitemShut {NoStop}%
\bibitem [{\citenamefont {Tai}\ and\ \citenamefont
  {Smalyukh}(2018)}]{tai_static_2018}%
  \BibitemOpen
  \bibfield  {author} {\bibinfo {author} {\bibfnamefont {J.-S.~B.}\
  \bibnamefont {Tai}}\ and\ \bibinfo {author} {\bibfnamefont {I.~I.}\
  \bibnamefont {Smalyukh}},\ }\href {\doibase 10.1103/PhysRevLett.121.187201}
  {\bibfield  {journal} {\bibinfo  {journal} {Phys. Rev. Lett.}\ }\textbf
  {\bibinfo {volume} {121}},\ \bibinfo {pages} {187201} (\bibinfo {year}
  {2018})}\BibitemShut {NoStop}%
\bibitem [{\citenamefont {Rybakov}\ \emph {et~al.}(2019)\citenamefont
  {Rybakov}, \citenamefont {Kiselev}, \citenamefont {Borisov}, \citenamefont
  {D{\"o}ring}, \citenamefont {Melcher},\ and\ \citenamefont
  {Bl{\"u}gel}}]{rybakov_magnetic_2019}%
  \BibitemOpen
  \bibfield  {author} {\bibinfo {author} {\bibfnamefont {F.~N.}\ \bibnamefont
  {Rybakov}}, \bibinfo {author} {\bibfnamefont {N.~S.}\ \bibnamefont
  {Kiselev}}, \bibinfo {author} {\bibfnamefont {A.~B.}\ \bibnamefont
  {Borisov}}, \bibinfo {author} {\bibfnamefont {L.}~\bibnamefont {D{\"o}ring}},
  \bibinfo {author} {\bibfnamefont {C.}~\bibnamefont {Melcher}}, \ and\
  \bibinfo {author} {\bibfnamefont {S.}~\bibnamefont {Bl{\"u}gel}},\ }\href
  {http://arxiv.org/abs/1904.00250} {\bibfield  {journal} {\bibinfo  {journal}
  {arXiv:1904.00250 [cond-mat, physics:nlin]}\ } (\bibinfo {year}
  {2019})}\BibitemShut {NoStop}%
\bibitem [{\citenamefont {Voinescu}\ \emph {et~al.}(2020)\citenamefont
  {Voinescu}, \citenamefont {Tai},\ and\ \citenamefont
  {Smalyukh}}]{voinescu_hopf_2020}%
  \BibitemOpen
  \bibfield  {author} {\bibinfo {author} {\bibfnamefont {R.}~\bibnamefont
  {Voinescu}}, \bibinfo {author} {\bibfnamefont {J.-S.~B.}\ \bibnamefont
  {Tai}}, \ and\ \bibinfo {author} {\bibfnamefont {I.~I.}\ \bibnamefont
  {Smalyukh}},\ }\href {\doibase 10.1103/PhysRevLett.125.057201} {\bibfield
  {journal} {\bibinfo  {journal} {Phys. Rev. Lett.}\ }\textbf {\bibinfo
  {volume} {125}},\ \bibinfo {pages} {057201} (\bibinfo {year}
  {2020})}\BibitemShut {NoStop}%
\bibitem [{\citenamefont {Kent}\ \emph {et~al.}(2021)\citenamefont {Kent},
  \citenamefont {Reynolds}, \citenamefont {Raftrey}, \citenamefont {Campbell},
  \citenamefont {Virasawmy}, \citenamefont {Dhuey}, \citenamefont {Chopdekar},
  \citenamefont {Hierro-Rodriguez}, \citenamefont {Sorrentino}, \citenamefont
  {Pereiro}, \citenamefont {Ferrer}, \citenamefont {Hellman}, \citenamefont
  {Sutcliffe},\ and\ \citenamefont {Fischer}}]{kent_creation_2021}%
  \BibitemOpen
  \bibfield  {author} {\bibinfo {author} {\bibfnamefont {N.}~\bibnamefont
  {Kent}}, \bibinfo {author} {\bibfnamefont {N.}~\bibnamefont {Reynolds}},
  \bibinfo {author} {\bibfnamefont {D.}~\bibnamefont {Raftrey}}, \bibinfo
  {author} {\bibfnamefont {I.~T.~G.}\ \bibnamefont {Campbell}}, \bibinfo
  {author} {\bibfnamefont {S.}~\bibnamefont {Virasawmy}}, \bibinfo {author}
  {\bibfnamefont {S.}~\bibnamefont {Dhuey}}, \bibinfo {author} {\bibfnamefont
  {R.~V.}\ \bibnamefont {Chopdekar}}, \bibinfo {author} {\bibfnamefont
  {A.}~\bibnamefont {Hierro-Rodriguez}}, \bibinfo {author} {\bibfnamefont
  {A.}~\bibnamefont {Sorrentino}}, \bibinfo {author} {\bibfnamefont
  {E.}~\bibnamefont {Pereiro}}, \bibinfo {author} {\bibfnamefont
  {S.}~\bibnamefont {Ferrer}}, \bibinfo {author} {\bibfnamefont
  {F.}~\bibnamefont {Hellman}}, \bibinfo {author} {\bibfnamefont
  {P.}~\bibnamefont {Sutcliffe}}, \ and\ \bibinfo {author} {\bibfnamefont
  {P.}~\bibnamefont {Fischer}},\ }\href {\doibase 10.1038/s41467-021-21846-5}
  {\bibfield  {journal} {\bibinfo  {journal} {Nat Commun}\ }\textbf {\bibinfo
  {volume} {12}},\ \bibinfo {pages} {1562} (\bibinfo {year}
  {2021})}\BibitemShut {NoStop}%
\bibitem [{\citenamefont {Whitehead}(1947)}]{whitehead_expression_1947}%
  \BibitemOpen
  \bibfield  {author} {\bibinfo {author} {\bibfnamefont {J.~H.~C.}\
  \bibnamefont {Whitehead}},\ }\href {http://www.pnas.org/content/33/5/117}
  {\bibfield  {journal} {\bibinfo  {journal} {PNAS}\ }\textbf {\bibinfo
  {volume} {33}},\ \bibinfo {pages} {117} (\bibinfo {year} {1947})}\BibitemShut
  {NoStop}%
\bibitem [{\citenamefont {Wilczek}\ and\ \citenamefont
  {Zee}(1983)}]{wilczek_linking_1983}%
  \BibitemOpen
  \bibfield  {author} {\bibinfo {author} {\bibfnamefont {F.}~\bibnamefont
  {Wilczek}}\ and\ \bibinfo {author} {\bibfnamefont {A.}~\bibnamefont {Zee}},\
  }\href {\doibase 10.1103/PhysRevLett.51.2250} {\bibfield  {journal} {\bibinfo
   {journal} {Phys. Rev. Lett.}\ }\textbf {\bibinfo {volume} {51}},\ \bibinfo
  {pages} {2250} (\bibinfo {year} {1983})}\BibitemShut {NoStop}%
\bibitem [{\citenamefont {Manton}\ and\ \citenamefont
  {Sutcliffe}(2004)}]{manton_topological_2004}%
  \BibitemOpen
  \bibfield  {author} {\bibinfo {author} {\bibfnamefont {N.}~\bibnamefont
  {Manton}}\ and\ \bibinfo {author} {\bibfnamefont {P.}~\bibnamefont
  {Sutcliffe}},\ }\href@noop {} {\emph {\bibinfo {title} {Topological
  {Solitons}}}}\ (\bibinfo  {publisher} {Cambridge University Press},\ \bibinfo
  {address} {Cambridge, England,},\ \bibinfo {year} {2004})\BibitemShut
  {NoStop}%
\bibitem [{\citenamefont {Liu}\ \emph {et~al.}(2020)\citenamefont {Liu},
  \citenamefont {Hou}, \citenamefont {Han},\ and\ \citenamefont
  {Zang}}]{liu_three_dimensional_2020}%
  \BibitemOpen
  \bibfield  {author} {\bibinfo {author} {\bibfnamefont {Y.}~\bibnamefont
  {Liu}}, \bibinfo {author} {\bibfnamefont {W.}~\bibnamefont {Hou}}, \bibinfo
  {author} {\bibfnamefont {X.}~\bibnamefont {Han}}, \ and\ \bibinfo {author}
  {\bibfnamefont {J.}~\bibnamefont {Zang}},\ }\href {\doibase
  10.1103/PhysRevLett.124.127204} {\bibfield  {journal} {\bibinfo  {journal}
  {Phys. Rev. Lett.}\ }\textbf {\bibinfo {volume} {124}},\ \bibinfo {pages}
  {127204} (\bibinfo {year} {2020})}\BibitemShut {NoStop}%
\bibitem [{\citenamefont {Wang}\ \emph {et~al.}(2019)\citenamefont {Wang},
  \citenamefont {Qaiumzadeh},\ and\ \citenamefont
  {Brataas}}]{wang_current_driven_2019}%
  \BibitemOpen
  \bibfield  {author} {\bibinfo {author} {\bibfnamefont {X.}~\bibnamefont
  {Wang}}, \bibinfo {author} {\bibfnamefont {A.}~\bibnamefont {Qaiumzadeh}}, \
  and\ \bibinfo {author} {\bibfnamefont {A.}~\bibnamefont {Brataas}},\ }\href
  {\doibase 10.1103/PhysRevLett.123.147203} {\bibfield  {journal} {\bibinfo
  {journal} {Phys. Rev. Lett.}\ }\textbf {\bibinfo {volume} {123}},\ \bibinfo
  {pages} {147203} (\bibinfo {year} {2019})}\BibitemShut {NoStop}%
\bibitem [{\citenamefont {Raftrey}\ and\ \citenamefont
  {Fischer}(2021)}]{raftrey_field-driven_2021}%
  \BibitemOpen
  \bibfield  {author} {\bibinfo {author} {\bibfnamefont {D.}~\bibnamefont
  {Raftrey}}\ and\ \bibinfo {author} {\bibfnamefont {P.}~\bibnamefont
  {Fischer}},\ }\href {\doibase 10.1103/PhysRevLett.127.257201} {\bibfield
  {journal} {\bibinfo  {journal} {Phys. Rev. Lett.}\ }\textbf {\bibinfo
  {volume} {127}},\ \bibinfo {pages} {257201} (\bibinfo {year}
  {2021})}\BibitemShut {NoStop}%
\bibitem [{\citenamefont {Bo}\ \emph {et~al.}(2021)\citenamefont {Bo},
  \citenamefont {Ji}, \citenamefont {Hu}, \citenamefont {Zhao}, \citenamefont
  {Li}, \citenamefont {Zhang},\ and\ \citenamefont {Zhang}}]{bo_spin_2021}%
  \BibitemOpen
  \bibfield  {author} {\bibinfo {author} {\bibfnamefont {L.}~\bibnamefont
  {Bo}}, \bibinfo {author} {\bibfnamefont {L.}~\bibnamefont {Ji}}, \bibinfo
  {author} {\bibfnamefont {C.}~\bibnamefont {Hu}}, \bibinfo {author}
  {\bibfnamefont {R.}~\bibnamefont {Zhao}}, \bibinfo {author} {\bibfnamefont
  {Y.}~\bibnamefont {Li}}, \bibinfo {author} {\bibfnamefont {J.}~\bibnamefont
  {Zhang}}, \ and\ \bibinfo {author} {\bibfnamefont {X.}~\bibnamefont
  {Zhang}},\ }\href {\doibase 10.1063/5.0072349} {\bibfield  {journal}
  {\bibinfo  {journal} {Appl. Phys. Lett.}\ }\textbf {\bibinfo {volume}
  {119}},\ \bibinfo {pages} {212408} (\bibinfo {year} {2021})}\BibitemShut
  {NoStop}%
\bibitem [{\citenamefont {Khodzhaev}\ and\ \citenamefont
  {Turgut}(2022)}]{khodzhaev_hopfion_2022}%
  \BibitemOpen
  \bibfield  {author} {\bibinfo {author} {\bibfnamefont {Z.}~\bibnamefont
  {Khodzhaev}}\ and\ \bibinfo {author} {\bibfnamefont {E.}~\bibnamefont
  {Turgut}},\ }\href {\doibase 10.1088/1361-648X/ac533d} {\bibfield  {journal}
  {\bibinfo  {journal} {J. Phys.: Condens. Matter}\ }\textbf {\bibinfo {volume}
  {34}},\ \bibinfo {pages} {225805} (\bibinfo {year} {2022})}\BibitemShut
  {NoStop}%
\bibitem [{\citenamefont {Li}\ \emph {et~al.}(2022)\citenamefont {Li},
  \citenamefont {Xia}, \citenamefont {Shen}, \citenamefont {Zhang},
  \citenamefont {Ezawa},\ and\ \citenamefont {Zhou}}]{li_mutual_2022}%
  \BibitemOpen
  \bibfield  {author} {\bibinfo {author} {\bibfnamefont {S.}~\bibnamefont
  {Li}}, \bibinfo {author} {\bibfnamefont {J.}~\bibnamefont {Xia}}, \bibinfo
  {author} {\bibfnamefont {L.}~\bibnamefont {Shen}}, \bibinfo {author}
  {\bibfnamefont {X.}~\bibnamefont {Zhang}}, \bibinfo {author} {\bibfnamefont
  {M.}~\bibnamefont {Ezawa}}, \ and\ \bibinfo {author} {\bibfnamefont
  {Y.}~\bibnamefont {Zhou}},\ }\href {\doibase 10.1103/PhysRevB.105.174407}
  {\bibfield  {journal} {\bibinfo  {journal} {Phys. Rev. B}\ }\textbf {\bibinfo
  {volume} {105}},\ \bibinfo {pages} {174407} (\bibinfo {year}
  {2022})}\BibitemShut {NoStop}%
\bibitem [{\citenamefont {G{\"o}bel}\ \emph {et~al.}(2020)\citenamefont
  {G{\"o}bel}, \citenamefont {Akosa}, \citenamefont {Tatara},\ and\
  \citenamefont {Mertig}}]{gobel_topological_2020}%
  \BibitemOpen
  \bibfield  {author} {\bibinfo {author} {\bibfnamefont {B.}~\bibnamefont
  {G{\"o}bel}}, \bibinfo {author} {\bibfnamefont {C.~A.}\ \bibnamefont
  {Akosa}}, \bibinfo {author} {\bibfnamefont {G.}~\bibnamefont {Tatara}}, \
  and\ \bibinfo {author} {\bibfnamefont {I.}~\bibnamefont {Mertig}},\ }\href
  {\doibase 10.1103/PhysRevResearch.2.013315} {\bibfield  {journal} {\bibinfo
  {journal} {Phys. Rev. Research}\ }\textbf {\bibinfo {volume} {2}},\ \bibinfo
  {pages} {013315} (\bibinfo {year} {2020})}\BibitemShut {NoStop}%
\bibitem [{\citenamefont {Pershoguba}\ \emph {et~al.}(2021)\citenamefont
  {Pershoguba}, \citenamefont {Andreoli},\ and\ \citenamefont
  {Zang}}]{pershoguba_electronic_2021}%
  \BibitemOpen
  \bibfield  {author} {\bibinfo {author} {\bibfnamefont {S.~S.}\ \bibnamefont
  {Pershoguba}}, \bibinfo {author} {\bibfnamefont {D.}~\bibnamefont
  {Andreoli}}, \ and\ \bibinfo {author} {\bibfnamefont {J.}~\bibnamefont
  {Zang}},\ }\href {\doibase 10.1103/PhysRevB.104.075102} {\bibfield  {journal}
  {\bibinfo  {journal} {Phys. Rev. B}\ }\textbf {\bibinfo {volume} {104}},\
  \bibinfo {pages} {075102} (\bibinfo {year} {2021})}\BibitemShut {NoStop}%
\bibitem [{sup()}]{supp}%
  \BibitemOpen
  \href@noop {} {}\bibinfo {note} {See Supplemental Materials for details of
  the derivations, numerical simulations, nonreciprocal dynamics of hopfion in
  chiral magnets, and movie of the nonreciprocal dynamics.}\BibitemShut {Stop}%
\bibitem [{\citenamefont {Auerbach}(1994)}]{auerbach_interacting_1994}%
  \BibitemOpen
  \bibfield  {author} {\bibinfo {author} {\bibfnamefont {A.}~\bibnamefont
  {Auerbach}},\ }\href {\doibase 10.1007/978-1-4612-0869-3} {\emph {\bibinfo
  {title} {Interacting {Electrons} and {Quantum} {Magnetism}}}},\ Graduate
  {Texts} in {Contemporary} {Physics}\ (\bibinfo  {publisher}
  {Springer-Verlag},\ \bibinfo {address} {New York},\ \bibinfo {year}
  {1994})\BibitemShut {NoStop}%
\bibitem [{\citenamefont {Tatara}\ \emph {et~al.}(2008)\citenamefont {Tatara},
  \citenamefont {Kohno},\ and\ \citenamefont
  {Shibata}}]{tatara_microscopic_2008}%
  \BibitemOpen
  \bibfield  {author} {\bibinfo {author} {\bibfnamefont {G.}~\bibnamefont
  {Tatara}}, \bibinfo {author} {\bibfnamefont {H.}~\bibnamefont {Kohno}}, \
  and\ \bibinfo {author} {\bibfnamefont {J.}~\bibnamefont {Shibata}},\ }\href
  {\doibase 10.1016/j.physrep.2008.07.003} {\bibfield  {journal} {\bibinfo
  {journal} {Physics Reports}\ }\textbf {\bibinfo {volume} {468}},\ \bibinfo
  {pages} {213} (\bibinfo {year} {2008})}\BibitemShut {NoStop}%
\bibitem [{\citenamefont {Watanabe}\ and\ \citenamefont
  {Yanase}(2018)}]{watanabe_group_theoretical_2018}%
  \BibitemOpen
  \bibfield  {author} {\bibinfo {author} {\bibfnamefont {H.}~\bibnamefont
  {Watanabe}}\ and\ \bibinfo {author} {\bibfnamefont {Y.}~\bibnamefont
  {Yanase}},\ }\href {\doibase 10.1103/PhysRevB.98.245129} {\bibfield
  {journal} {\bibinfo  {journal} {Phys. Rev. B}\ }\textbf {\bibinfo {volume}
  {98}},\ \bibinfo {pages} {245129} (\bibinfo {year} {2018})}\BibitemShut
  {NoStop}%
\bibitem [{\citenamefont {Hayami}\ \emph {et~al.}(2018)\citenamefont {Hayami},
  \citenamefont {Yatsushiro}, \citenamefont {Yanagi},\ and\ \citenamefont
  {Kusunose}}]{hayami_classification_2018}%
  \BibitemOpen
  \bibfield  {author} {\bibinfo {author} {\bibfnamefont {S.}~\bibnamefont
  {Hayami}}, \bibinfo {author} {\bibfnamefont {M.}~\bibnamefont {Yatsushiro}},
  \bibinfo {author} {\bibfnamefont {Y.}~\bibnamefont {Yanagi}}, \ and\ \bibinfo
  {author} {\bibfnamefont {H.}~\bibnamefont {Kusunose}},\ }\href {\doibase
  10.1103/PhysRevB.98.165110} {\bibfield  {journal} {\bibinfo  {journal} {Phys.
  Rev. B}\ }\textbf {\bibinfo {volume} {98}},\ \bibinfo {pages} {165110}
  (\bibinfo {year} {2018})}\BibitemShut {NoStop}%
\bibitem [{\citenamefont {Papanicolaou}\ and\ \citenamefont
  {Tomaras}(1991)}]{papanicolaou_dynamics_1991}%
  \BibitemOpen
  \bibfield  {author} {\bibinfo {author} {\bibfnamefont {N.}~\bibnamefont
  {Papanicolaou}}\ and\ \bibinfo {author} {\bibfnamefont {T.~N.}\ \bibnamefont
  {Tomaras}},\ }\href {\doibase 10.1016/0550-3213(91)90410-Y} {\bibfield
  {journal} {\bibinfo  {journal} {Nuclear Physics B}\ }\textbf {\bibinfo
  {volume} {360}},\ \bibinfo {pages} {425} (\bibinfo {year}
  {1991})}\BibitemShut {NoStop}%
\bibitem [{\citenamefont {Guslienko}(2008)}]{guslienko_magnetic_2008}%
  \BibitemOpen
  \bibfield  {author} {\bibinfo {author} {\bibfnamefont {K.~Y.}\ \bibnamefont
  {Guslienko}},\ }\href {\doibase doi:10.1166/jnn.2008.003} {\bibfield
  {journal} {\bibinfo  {journal} {J. Nanosci. Nanotechnol.}\ }\textbf {\bibinfo
  {volume} {8}},\ \bibinfo {pages} {2745} (\bibinfo {year} {2008})}\BibitemShut
  {NoStop}%
\bibitem [{\citenamefont {Tretiakov}\ \emph {et~al.}(2008)\citenamefont
  {Tretiakov}, \citenamefont {Clarke}, \citenamefont {Chern}, \citenamefont
  {Bazaliy},\ and\ \citenamefont {Tchernyshyov}}]{tretiakov_dynamics_2008}%
  \BibitemOpen
  \bibfield  {author} {\bibinfo {author} {\bibfnamefont {O.~A.}\ \bibnamefont
  {Tretiakov}}, \bibinfo {author} {\bibfnamefont {D.}~\bibnamefont {Clarke}},
  \bibinfo {author} {\bibfnamefont {G.-W.}\ \bibnamefont {Chern}}, \bibinfo
  {author} {\bibfnamefont {Y.~B.}\ \bibnamefont {Bazaliy}}, \ and\ \bibinfo
  {author} {\bibfnamefont {O.}~\bibnamefont {Tchernyshyov}},\ }\href {\doibase
  10.1103/PhysRevLett.100.127204} {\bibfield  {journal} {\bibinfo  {journal}
  {Phys. Rev. Lett.}\ }\textbf {\bibinfo {volume} {100}},\ \bibinfo {pages}
  {127204} (\bibinfo {year} {2008})}\BibitemShut {NoStop}%
\bibitem [{\citenamefont {Everschor-Sitte}\ and\ \citenamefont
  {Sitte}(2014)}]{everschor_sitte_real_space_2014}%
  \BibitemOpen
  \bibfield  {author} {\bibinfo {author} {\bibfnamefont {K.}~\bibnamefont
  {Everschor-Sitte}}\ and\ \bibinfo {author} {\bibfnamefont {M.}~\bibnamefont
  {Sitte}},\ }\href {\doibase 10.1063/1.4870695} {\bibfield  {journal}
  {\bibinfo  {journal} {Journal of Applied Physics}\ }\textbf {\bibinfo
  {volume} {115}},\ \bibinfo {pages} {172602} (\bibinfo {year}
  {2014})}\BibitemShut {NoStop}%
\bibitem [{\citenamefont {Psaroudaki}\ and\ \citenamefont
  {Loss}(2018)}]{psaroudaki_skyrmions_2018}%
  \BibitemOpen
  \bibfield  {author} {\bibinfo {author} {\bibfnamefont {C.}~\bibnamefont
  {Psaroudaki}}\ and\ \bibinfo {author} {\bibfnamefont {D.}~\bibnamefont
  {Loss}},\ }\href {\doibase 10.1103/PhysRevLett.120.237203} {\bibfield
  {journal} {\bibinfo  {journal} {Phys. Rev. Lett.}\ }\textbf {\bibinfo
  {volume} {120}},\ \bibinfo {pages} {237203} (\bibinfo {year}
  {2018})}\BibitemShut {NoStop}%
\bibitem [{\citenamefont {Cheong}\ and\ \citenamefont
  {Xu}(2022)}]{cheong_magnetic_2022}%
  \BibitemOpen
  \bibfield  {author} {\bibinfo {author} {\bibfnamefont {S.-W.}\ \bibnamefont
  {Cheong}}\ and\ \bibinfo {author} {\bibfnamefont {X.}~\bibnamefont {Xu}},\
  }\href {\doibase 10.1038/s41535-022-00447-5} {\bibfield  {journal} {\bibinfo
  {journal} {npj Quantum Mater.}\ }\textbf {\bibinfo {volume} {7}},\ \bibinfo
  {pages} {1} (\bibinfo {year} {2022})}\BibitemShut {NoStop}%
\bibitem [{\citenamefont
  {Slonczewski}(1996)}]{slonczewski_current-driven_1996}%
  \BibitemOpen
  \bibfield  {author} {\bibinfo {author} {\bibfnamefont {J.~C.}\ \bibnamefont
  {Slonczewski}},\ }\href {\doibase 10.1016/0304-8853(96)00062-5} {\bibfield
  {journal} {\bibinfo  {journal} {Journal of Magnetism and Magnetic Materials}\
  }\textbf {\bibinfo {volume} {159}},\ \bibinfo {pages} {L1} (\bibinfo {year}
  {1996})}\BibitemShut {NoStop}%
\bibitem [{\citenamefont {Berger}(1996)}]{berger_emission_1996}%
  \BibitemOpen
  \bibfield  {author} {\bibinfo {author} {\bibfnamefont {L.}~\bibnamefont
  {Berger}},\ }\href {\doibase 10.1103/PhysRevB.54.9353} {\bibfield  {journal}
  {\bibinfo  {journal} {Phys. Rev. B}\ }\textbf {\bibinfo {volume} {54}},\
  \bibinfo {pages} {9353} (\bibinfo {year} {1996})}\BibitemShut {NoStop}%
\bibitem [{\citenamefont {Zhang}\ and\ \citenamefont
  {Li}(2004)}]{zhang_roles_2004}%
  \BibitemOpen
  \bibfield  {author} {\bibinfo {author} {\bibfnamefont {S.}~\bibnamefont
  {Zhang}}\ and\ \bibinfo {author} {\bibfnamefont {Z.}~\bibnamefont {Li}},\
  }\href {\doibase 10.1103/PhysRevLett.93.127204} {\bibfield  {journal}
  {\bibinfo  {journal} {Phys. Rev. Lett.}\ }\textbf {\bibinfo {volume} {93}},\
  \bibinfo {pages} {127204} (\bibinfo {year} {2004})}\BibitemShut {NoStop}%
\bibitem [{\citenamefont {Thiele}(1973)}]{thiele_steady-state_1973}%
  \BibitemOpen
  \bibfield  {author} {\bibinfo {author} {\bibfnamefont {A.~A.}\ \bibnamefont
  {Thiele}},\ }\href {\doibase 10.1103/PhysRevLett.30.230} {\bibfield
  {journal} {\bibinfo  {journal} {Phys. Rev. Lett.}\ }\textbf {\bibinfo
  {volume} {30}},\ \bibinfo {pages} {230} (\bibinfo {year} {1973})}\BibitemShut
  {NoStop}%
\bibitem [{\citenamefont {Ye}\ \emph {et~al.}(1999)\citenamefont {Ye},
  \citenamefont {Kim}, \citenamefont {Millis}, \citenamefont {Shraiman},
  \citenamefont {Majumdar},\ and\ \citenamefont {Tešanović}}]{ye_berry_1999}%
  \BibitemOpen
  \bibfield  {author} {\bibinfo {author} {\bibfnamefont {J.}~\bibnamefont
  {Ye}}, \bibinfo {author} {\bibfnamefont {Y.~B.}\ \bibnamefont {Kim}},
  \bibinfo {author} {\bibfnamefont {A.~J.}\ \bibnamefont {Millis}}, \bibinfo
  {author} {\bibfnamefont {B.~I.}\ \bibnamefont {Shraiman}}, \bibinfo {author}
  {\bibfnamefont {P.}~\bibnamefont {Majumdar}}, \ and\ \bibinfo {author}
  {\bibfnamefont {Z.}~\bibnamefont {Tešanović}},\ }\href {\doibase
  10.1103/PhysRevLett.83.3737} {\bibfield  {journal} {\bibinfo  {journal}
  {Phys. Rev. Lett.}\ }\textbf {\bibinfo {volume} {83}},\ \bibinfo {pages}
  {3737} (\bibinfo {year} {1999})}\BibitemShut {NoStop}%
\bibitem [{\citenamefont {Bruno}\ \emph {et~al.}(2004)\citenamefont {Bruno},
  \citenamefont {Dugaev},\ and\ \citenamefont
  {Taillefumier}}]{bruno_topological_2004}%
  \BibitemOpen
  \bibfield  {author} {\bibinfo {author} {\bibfnamefont {P.}~\bibnamefont
  {Bruno}}, \bibinfo {author} {\bibfnamefont {V.~K.}\ \bibnamefont {Dugaev}}, \
  and\ \bibinfo {author} {\bibfnamefont {M.}~\bibnamefont {Taillefumier}},\
  }\href {\doibase 10.1103/PhysRevLett.93.096806} {\bibfield  {journal}
  {\bibinfo  {journal} {Phys. Rev. Lett.}\ }\textbf {\bibinfo {volume} {93}},\
  \bibinfo {pages} {096806} (\bibinfo {year} {2004})}\BibitemShut {NoStop}%
\bibitem [{\citenamefont {Isobe}\ \emph {et~al.}(2020)\citenamefont {Isobe},
  \citenamefont {Xu},\ and\ \citenamefont {Fu}}]{isobe_high-frequency_2020}%
  \BibitemOpen
  \bibfield  {author} {\bibinfo {author} {\bibfnamefont {H.}~\bibnamefont
  {Isobe}}, \bibinfo {author} {\bibfnamefont {S.-Y.}\ \bibnamefont {Xu}}, \
  and\ \bibinfo {author} {\bibfnamefont {L.}~\bibnamefont {Fu}},\ }\href
  {\doibase 10.1126/sciadv.aay2497} {\bibfield  {journal} {\bibinfo  {journal}
  {Science Advances}\ }\textbf {\bibinfo {volume} {6}},\ \bibinfo {pages}
  {eaay2497} (\bibinfo {year} {2020})}\BibitemShut {NoStop}%
\bibitem [{\citenamefont {Ishizuka}\ and\ \citenamefont
  {Nagaosa}(2020)}]{ishizuka_anomalous_2020}%
  \BibitemOpen
  \bibfield  {author} {\bibinfo {author} {\bibfnamefont {H.}~\bibnamefont
  {Ishizuka}}\ and\ \bibinfo {author} {\bibfnamefont {N.}~\bibnamefont
  {Nagaosa}},\ }\href {\doibase 10.1038/s41467-020-16751-2} {\bibfield
  {journal} {\bibinfo  {journal} {Nat Commun}\ }\textbf {\bibinfo {volume}
  {11}},\ \bibinfo {pages} {2986} (\bibinfo {year} {2020})}\BibitemShut
  {NoStop}%
\bibitem [{\citenamefont {Isobe}\ and\ \citenamefont
  {Nagaosa}(2022)}]{isobe_toroidal_2022}%
  \BibitemOpen
  \bibfield  {author} {\bibinfo {author} {\bibfnamefont {H.}~\bibnamefont
  {Isobe}}\ and\ \bibinfo {author} {\bibfnamefont {N.}~\bibnamefont
  {Nagaosa}},\ }\href {\doibase 10.7566/JPSJ.91.115001} {\bibfield  {journal}
  {\bibinfo  {journal} {J. Phys. Soc. Jpn.}\ }\textbf {\bibinfo {volume}
  {91}},\ \bibinfo {pages} {115001} (\bibinfo {year} {2022})}\BibitemShut
  {NoStop}%
\bibitem [{\citenamefont {Volovik}\ and\ \citenamefont
  {Mineev}(1977)}]{volovik_1977}%
  \BibitemOpen
  \bibfield  {author} {\bibinfo {author} {\bibfnamefont {G.}~\bibnamefont
  {Volovik}}\ and\ \bibinfo {author} {\bibfnamefont {V.}~\bibnamefont
  {Mineev}},\ }\href@noop {} {\bibfield  {journal} {\bibinfo  {journal} {Sov.
  Phys. JETP}\ }\textbf {\bibinfo {volume} {46}},\ \bibinfo {pages} {401}
  (\bibinfo {year} {1977})}\BibitemShut {NoStop}%
\bibitem [{\citenamefont {Babaev}(2002)}]{babaev_dual_2002}%
  \BibitemOpen
  \bibfield  {author} {\bibinfo {author} {\bibfnamefont {E.}~\bibnamefont
  {Babaev}},\ }\href {\doibase 10.1103/PhysRevLett.88.177002} {\bibfield
  {journal} {\bibinfo  {journal} {Phys. Rev. Lett.}\ }\textbf {\bibinfo
  {volume} {88}},\ \bibinfo {pages} {177002} (\bibinfo {year}
  {2002})}\BibitemShut {NoStop}%
\bibitem [{\citenamefont {Kawaguchi}\ \emph {et~al.}(2008)\citenamefont
  {Kawaguchi}, \citenamefont {Nitta},\ and\ \citenamefont
  {Ueda}}]{kawaguchi_knots_2008}%
  \BibitemOpen
  \bibfield  {author} {\bibinfo {author} {\bibfnamefont {Y.}~\bibnamefont
  {Kawaguchi}}, \bibinfo {author} {\bibfnamefont {M.}~\bibnamefont {Nitta}}, \
  and\ \bibinfo {author} {\bibfnamefont {M.}~\bibnamefont {Ueda}},\ }\href
  {\doibase 10.1103/PhysRevLett.100.180403} {\bibfield  {journal} {\bibinfo
  {journal} {Phys. Rev. Lett.}\ }\textbf {\bibinfo {volume} {100}},\ \bibinfo
  {pages} {180403} (\bibinfo {year} {2008})}\BibitemShut {NoStop}%
\bibitem [{\citenamefont {Ackerman}\ and\ \citenamefont
  {Smalyukh}(2017)}]{ackerman_static_2017}%
  \BibitemOpen
  \bibfield  {author} {\bibinfo {author} {\bibfnamefont {P.~J.}\ \bibnamefont
  {Ackerman}}\ and\ \bibinfo {author} {\bibfnamefont {I.~I.}\ \bibnamefont
  {Smalyukh}},\ }\href {\doibase 10.1038/nmat4826} {\bibfield  {journal}
  {\bibinfo  {journal} {Nat. Mater.}\ }\textbf {\bibinfo {volume} {16}},\
  \bibinfo {pages} {426} (\bibinfo {year} {2017})}\BibitemShut {NoStop}%
\end{thebibliography}
%

\end{document}